\documentclass[11pt,final, sort&compress]{elsarticle}

\usepackage{amssymb}
\usepackage{amsmath}
\usepackage{graphicx}
\usepackage{tabularx}
\usepackage{cite}
\usepackage{color}
\usepackage{fullpage}
\usepackage{subfigure}
\usepackage{subfigmat}
\usepackage{etoolbox}
\usepackage{url}
\usepackage{float}
\usepackage{bm}

\usepackage{xcolor}
\definecolor{grayBlue_Matlab}{rgb}{0 0.45 0.75}
\definecolor{Gray_Matlab}{rgb}{0.6 0.6 0.6}
\definecolor{BlackGray_Matlab}{rgb}{0.33 0.33 0.33}

\usepackage{siunitx}
\usepackage{multicol}
\usepackage{nomencl}
\usepackage{mdframed}
\usepackage[section]{placeins}

\renewcommand{\nomname}{NOMENCLATURE}

\newcommand{\nomunit}[1]{%
\renewcommand{\nomentryend}{\hspace*{\fill}#1}}

\RequirePackage{ifthen}
\renewcommand{\nomgroup}[1]{%
  \ifthenelse{\equal{#1}{A}}{\item[\textbf{Latin Symbols:}]}{%
    \ifthenelse{\equal{#1}{I}}{\item[\textbf{Subscripts:}]}{%
        \ifthenelse{\equal{#1}{S}}{\item[\textbf{Superscripts:}]}{%
          \ifthenelse{\equal{#1}{M}}{\item[\textbf{Ohers:}]}{%
          \ifthenelse{\equal{#1}{Z}}{\item[\textbf{Dimensionsless Numbers:}]}{%
            \ifthenelse{\equal{#1}{G}}{\item[\textbf{Greek Symbols:}]}{}}}}}}}

\patchcmd{\thenomenclature}
 {\section*{\nomname}}
 {\begin{multicols}{2}[\section*{\nomname}]}
 {}{}
\appto\endthenomenclature{\end{multicols}}
\journal{Atomization and Sprays}

\begin{document}
\begin{frontmatter}

\title{Large eddy simulations of cavitating flow in a step nozzle \\ with injection into gas}

\author{Theresa Trummler}
\ead{theresa.trummler@tum.de}
\author{Daniel Rahn}
\author{Steffen J. Schmidt}
\author{Nikolaus A. Adams}

\address{
    Chair of Aerodynamics and Fluid Mechanics, Technical University of Munich \\ 
    Boltzmannstr. 15, 85748 Garching bei M\"unchen, Germany}

\begin{abstract}

    We present results of Large Eddy Simulations of a cavitating nozzle flow and injection into gas, investigating the interactions of cavitation in the nozzle, primary jet break-up, mass-flow rates and gas entrainment. During strong cavitation, detached vapor structures can reach the nozzle outlet, leading to partial entrainment  of gas from the outflow region into the nozzle. The gas entrainment can affect cavitation dynamics, mass-flow rates  and jet break-up. Moreover, the implosion of detached vapor structures induces pressure peaks that on the one hand amplify turbulent fluctuations and subsequently can enhance jet break-up and on the other hand can damage walls in the proximity and thus lead to cavitation erosion. 

	Our numerical setup is based on a reference experiment, in which liquid water is discharged into ambient air through a step nozzle. The cavitating liquid and the non-condensable gas phase are modeled with a barotropic homogeneous mixture model while for the numerical model a high-order implicit Large Eddy approach is employed. Full compressibility of all components is taken into account, enabling us to capture the effects of collapsing vapor structures. 

	Two operating points covering different cavitation regimes and jet characteristics are investigated. Special emphasis is placed on studying the effects of cavitation on the mass flow and the jet as well as the impact of partial gas entrainment.  Therefore, frequency analyses of the recorded time-resolved signals are performed. Furthermore, the dynamics and intensities of imploding vapor structures are assessed. 

\end{abstract}

\begin{keyword}
    cavitating nozzle flows, primary jet breakup, interaction cavitation mass flow, frequency analysis, gas entrainment, collapse dynamics
\end{keyword}

\end{frontmatter}
\makenomenclature
\section{Introduction}
\label{sec:introduction}

In pressure atomizers or injection systems the working fluid is strongly accelerated before it is discharged. This acceleration results in a local decrease of the static pressure, in some cases even below vapor pressure, which leads to cavitation. Liquid-embedded vapor structures are subsequently transported into regions of higher pressure where they collapse and emit intense shock waves. Cavitation in injection systems can have desired and undesired effects. The most beneficial one is the promotion of primary jet break up and fuel atomization \citep{bergwerk1959flow, reitz1982mechanism}. Since spray quality is one of the key parameters for combustion efficiency and reduction of pollutants, this can become a central aspect to fulfill future emission standards. Adverse effects are reduction of the mass flow and cavitation erosion. Cavitation can strongly interact with the mass flow, since the generated vapor reduces the effective cross section and the discharge coefficient, as already observed by \citet{bergwerk1959flow, nurick1976orifice}. This can have a large impact, since the amount of injected fuel is crucial for internal combustion processes and the efficiency. 

The implosion of detached vapor structures leads to the emission of shock waves and induces extremely high pressure peaks. Collapses near solid walls induce high surface loads, which can lead to material erosion \citep{philipp1998cavitation} and finally to severe material damage and even to injector failure \citep{asi2006failure}. 
In conclusion, it is vital to improve the understanding of cavitation and its consequences in order to optimize spray quality and the durability of injector components, and also to ensure a sufficient injection of fuel. 

Experimental investigations in real injectors are very challenging due to the high injection pressures and small dimensions. Additionally, the highly transient flow field and the lack of optical access make measurements difficult. To investigate the cavitation dynamics and the growth and subsequent collapse of vapor structures, optical access to the nozzle is required. Thus, most investigations of cavitation dynamics were performed in large scale models with acrylic glass or Lucite. The cavitation dynamics of internal nozzle flows was investigated experimentally by \citet{Sato:2002vv, Saito:2003us, Sugimoto:2009, Stanley:2014id}, while cavitating nozzle flow and jet were simultaneously analyzed in e.g.\citet{Sou:2007jd, Sou:2014hja, Pratama:2013vj, Stanley:2011gr}, providing a deeper insight into the interaction of cavitation and jet break-up.

For strong cavitation vapor structures reach the nozzle outlet and lead to the entrainment of gas into the nozzle. In the experiments a distinction between gas and vapor is hardly possible because of similar physical properties. Only recently \citet{duke2016x} presented an application of x-ray fluorescence in which vapor and gas in a cavitating nozzle were measured simultaneously. However, to the knowledge of the authors, there are no experimental references where partial gas-entrainment in the nozzle and its effects are assessed. Consequently, numerical simulations are essential to gain further insight into the effects of partial gas entrainment into the nozzle.

To assess the correlation of cavitation dynamics and the occurrence of collapse events, optical access must be available, and at the same time material impact must be detectable, ideally time resolved. The viable injection pressures are limited in large scale models with optical access and thus can not be increased to a level comparable to pressures in injector components. Consequently, these experiments mostly do not show visible material wear and do not provide collapse data of the detached vapor pockets. For cavitating flow around a wedge \citet{Petkovsek:2013ii} simultaneously monitored the cavitation dynamics with high speed cameras and visualized the material damage by using aluminum foil. To the knowledge of the authors, there are no such studies for cavitating nozzle flows and thus numerical simulations can provide supplementary data. 

In the last decades Computational Fluid Dynamics (CFD) have become a complementary approach for analyzing cavitation phenomena. Early numerical studies were performed with incompressible Reynolds-Averaged Navier Stokes methods \citep[e.g.][]{giannadakis2008modelling}. Later, more advanced methods such as time resolved compressible flow solvers where developed, which can capture the dynamics and the emitted shock-waves and thus able to predict cavitation erosion \citep{mihatsch2015cavitation, Beban:2017vo}. These simulations also provide detailed information on the temporal and spatial occurrence of collapses and their intensity.
Large Eddy Simulations (LES) have been established as an appropriate tool to obtain time-resolved data and to capture turbulent effects. By providing time-resolved, three-dimensional flow field data they allow for detailed analysis and can contribute to a better understanding of the underlying physical mechanisms. \citet{Egerer:2014wu} performed fully-compressible LES of a flow with developing cavitation and with inertia driven strong cavitation in a generic nozzle geometry. They proved the ability of the homogeneous mixture model, in combination with a cavitation model based on thermodynamic equilibrium, to treat appropriately both cavitation regimes. Additionally, \citet{he2016experimental} and \citet{Koukouvinis:2016boa} performed LES of incipient cavitation for submerged nozzle flows and further confirmed the applicability of LES to cavitation analysis by validation with experimental data. \citet{Koukouvinis:2016boa} included a comparison of various cavitation models in combination with a homogeneous mixture approach and demonstrated that for developing cavitation occurring in large scale step nozzles the barotropic equilibrium model leads to comparable cavitation development and volume fraction distribution as with relaxation models. \citet{Sou:2014hja} employed a different method, including an Eulerian-Lagrangian Bubble Tracking Method (BTM) in their LES and a cavitation model based on the Rayleigh-Plesset equation for incipient cavitation.

Full-configuration simulations including the cavitating nozzle flow and the jet are best suited to gain a deeper insight into the mutual interaction of cavitation and spray formation. For this reason, \citet{Orley:2015kt} extended the model utilized in \citet{Egerer:2014wu} by an additional non-condensable gas component and performed LES of a reference experiment. \citet{Edelbauer:2017fx} performed incompressible LES of a similar configuration, using an approach which combines the Euler-Euler modeling for the cavitating liquid with a Volume of Fluid Method for the spray at the liquid-gas interface. Additionally, \citet{orley2016large} performed a multi-component compressible LES of flows through injector components including needle movement and discharge of the fuel into a gaseous ambient. Nevertheless, there are still various interactions and mechanisms not yet fully understood, such as the effect of cavitation dynamics and especially of the gas entrainment on the mass flow and the formation of condensation shocks \citep{budich2018numerical}. 

In this paper, we carry out compressible LES of cavitating nozzle flows emanating into ambient air. Our numerical setup is based on reference experiments performed by \citet{Sou:2014hja} and \citet{Bicer:2015cp}. The non-dimensional numbers in the experiment - the Reynolds number and the Cavitation number - are in the same range as for typical technical applications. Two operating points covering different cavitation regimes are investigated. The employed cavitation model has already been successfully validated against reference experiments \citep{Orley:2015kt}. Moreover, past investigations using the same modeling approach and featuring significantly higher accelerations and velocities \citep{Egerer:2014wu, orley2016large} in smaller geometries, have demonstrated the feasibility of the single fluid approach. The thermodynamic model is embedded in a density-based fully compressible flow solver with a higher order implicit LES approach for narrow stencils \citep{Egerer:2016it}. For application of the multi-component model with an additional gas phase, we have extended the LES approach by \citet{Egerer:2016it}. Additionally, a thermodynamically consistent coupling within the numerical model is introduced to avoid unphysical oscillations of the pressure. The utilized multi-component model enables us to investigate the interactions between cavitation, mass-flow and jet characteristics within a single simulation of the cavitating nozzle flow and the jet. The fully compressible approach allows us to capture effects of collapsing vapor structures and to quantify their impact on turbulent fluctuations and, subsequently, the primary jet break-up. 

The paper is structured as follows. The numerical method and thermodynamic model are discussed in Section~\ref{sec:mathe_phys_model}. In Section~\ref{sec:numerical_setup} the numerical setup is presented. Our main results are discussed in Section~\ref{sec:results}, where we first compare them against experimental data, analyze the cavitation dynamics, investigate the interaction of cavitation and the mass-flow, perform frequency analysis, assess the dynamics and occurrence of collapse events and discuss the effects of cavitation on the jet characteristics. 
At the end of the paper the main conclusions are summarized in Section~\ref{sec:conclusion}.

\section{Mathematical and Physical Model}
\label{sec:mathe_phys_model}

In this section the mathematical and physical model are described. The governing equations are the Navier-Stokes equations and the thermodynamic modeling uses a homogeneous mixture model with a barotropic equation of state (EOS). The thermodynamic model is embedded in an implicit LES approach \citep{Egerer:2016it}, adapted for multiphase flows. In this paper an adapted sensor functional is presented and a physically consistent coupling for multi-components is introduced. 

\subsection{Governing Equations}
\label{subsec:gov_eq}
The governing equations are the fully compressible Navier-Stokes equations, written here in conservative form:
  \begin{equation}
     \partial_{t}\boldsymbol{U}+\nabla \cdot [ \boldsymbol{C}(\boldsymbol{U})+\boldsymbol{S}(\boldsymbol{U})]=0, 
     \label{eq:NS}
  \end{equation}
where $\boldsymbol{U}=[\rho , \, \rho u_{1}, \, \rho u_{2} , \, \rho u_{3} , \, \rho \xi ]$ is the vector of the conservative variables density $\rho$, momentum $ \rho u_{i}$ and gas density $\rho \xi $. The energy equation can be omitted due to the barotropic modeling of the cavitating liquid together with an isothermal modeling of the gas phase. The omission of the energy equation implies that heat-transfer is neglected. The numerical flux vector is divided into a convective part 
  \begin{equation}
     \boldsymbol{C}_{i}(\boldsymbol{U})=u_{i}\boldsymbol{U}
     \label{eq:C}
  \end{equation}
and a part $\boldsymbol{S}$ containing the surface stresses
  \begin{equation}
     \boldsymbol{S}_{i}(\boldsymbol{U})=[0,\delta_{i1}p-\tau_{i1}, \delta_{i2}p-\tau_{i2}, \delta_{i3}p-\tau_{i3},  0 ] \,,
     \label{eq:S}
  \end{equation}
where $\delta_{ij}$ is the Kronecker-Delta, $p$ the static pressure and $\boldsymbol{\tau}$ stands for the viscous stress tensor 
  \begin{equation}
    \tau_{ij}=\mu (\partial_{j}u_{i}+\partial_{i}u_{j}+\frac{2}{3}\delta_{ij}\partial_{k}u_{k}) \,,
    \label{eq:tau}
  \end{equation}
with the dynamic viscosity of the fluid $\mu$. The numerical evaluation of the fluxes at cell faces is discussed in Subsection~\ref{subsec:implicit_LES}. 

\subsection{Thermodynamic Model}
\label{subsec:thermo}
We employ the multi-component homogeneous mixture model proposed by \citet{Orley:2015kt}, in which cavitating liquid ($LV$) and non-condensable gas ($G$) are described by a single mixture fluid, defined by the volume averaged density inside a computational cell 
  \begin{equation}
     \rho=\sum_{\varphi} \beta_{\varphi} \rho_{\varphi} \,.
     \label{eq:beta}
  \end{equation}
$\beta_{\varphi}$ denotes the volume fraction and $\rho_{\varphi}$ the density of each component $\varphi= \{ LV, G\}$. This single fluid approach implies that within a computational cell all phases have the same velocity, temperature and pressure. A coupled equation of state (EOS) provides the pressure as a function of the mean density $ p = p(\rho)$ and is obtained by expressing the densities $\rho_{\varphi}$ using the corresponding thermodynamic relations.

The cavitating water is described with an isentropic EOS, derived by integration of the speed of sound c 
  \begin{equation}
     c=\sqrt{\dfrac{\partial p}{\partial \rho}\bigg|_{s=const.}} 
     \label{eq:c0}
  \end{equation}
which leads to 
  \begin{equation}
   \rho_{LV}=\rho_\mathrm{sat,L}+(p-p_\mathrm{sat}) / c^2\,,
  \end{equation}
where $\rho_\mathrm{sat,L}$ is the saturation density for liquid water and $p_\mathrm{sat}$ the saturation pressure. For water at the reference temperature $\mathrm{Temp}=\SI{293.15}{K}$ the values are $p_\mathrm{sat}=\SI{2340}{Pa}$ and $\rho_\mathrm{sat,L}=\SI{998.1618}{kg/m^3}$. Phase change is modeled assuming local thermodynamic equilibrium. For $p>p_\mathrm{sat}$, there is purely liquid water and the speed of sound $c=\SI{1482.35}{m/s}$. For $p<p_\mathrm{sat}$, there is a liquid vapor mixture. In the two phase region the structure of the EOS remains the same, but the value of the speed of sound is adapted to the properties in the two phase region. \citet{franc2004fundamentals} derived a formula for the speed of sound in the two phase region. They obtain for water and a vapor content of $\alpha=50\%$ an  equilibrium speed of sound considering phase change effects of  $c_{eq}=0.08\,\si{m/s}$.  If phase change is neglected, the frozen speed of sound is  derived $c_{frozen}=3\,\si{m/s}$. According to \citet{brennen1995cavitation} the speed of sound considering phase change is between these two bounds. We thus follow \citet{Orley:2015kt} and consider an average speed of sound of $c=\SI{1}{m/s}$. Moreover, since the speed of sound in the two phase region is significantly smaller than both the bulk velocity  and the interface velocity, and the flow thus is supersonic, the precise value of the speed of sound in the two phase region is not critical. 

Based on the density of the liquid vapor mixture we calculate the volume fraction of the vapor $\alpha$ with
  \begin{equation}
     \alpha=\frac{\rho_\mathrm{sat,L}-\rho_{LV}}{\rho_\mathrm{sat,L}-\rho_{\mathrm{sat,V}}},\; \;\; \mathrm{if} \;\;\rho_{LV} < \rho_\mathrm{sat,L}
  \end{equation}
where $\rho_{\mathrm{sat,V}}= \SI{0.017214 }{kg/m^3}$ denotes the saturation density of the vapor. 

The non-condensable gas-phase is modeled as an isothermal ideal gas obeying the ideal gas law 
  \begin{equation}
     \rho_{G}= \frac{p}{\mathrm{R}\cdot\mathrm{Temp}} . 
     \label{eq:ideale_gasgl}
  \end{equation}
$R$ is the specific gas constant with $R = \SI{287.06}{J/kg\cdot K}$ for air and temperature $\mathrm{Temp}=\SI{293.15}{K}$.  

Since the dynamic viscosities of vapor and liquid differ by several orders of magnitude, it is crucial to correctly compute the viscosity of the mixture fluid. We first estimate the dynamic viscosity of the liquid-vapor mixture and later combine it with the gas viscosity to obtain a mixture viscosity for all three components. For the dynamic viscosity of the liquid-vapor mixture we follow \citet{Beattie:1982ut} 
  \begin{equation}
     \mu_{LV}=\alpha\cdot\mu_{V}+(1-\alpha)(1+\frac{5}{2}\alpha)\mu_{L} \,,
     \label{eq:muLiqMix}
  \end{equation}
assuming that if only a small amount of vapor is present it will be distributed in the form of small vapor bubbles which behave like rigid particles. This approach has already been successfully used in similar applications \citep{Egerer:2014wu, Hickel:2014wz}. The viscosity of the mixture is then obtained by a linear blending with the volume fractions
  \begin{equation}
     \mu=\beta_{G}\cdot\mu_{G} + (1-\beta_{G})\cdot\mu_{LV} \,.
     \label{eq:muLiqMix}
  \end{equation}
We adopt the following viscosity values: $\mu_{L} = \SI{1.002e-3}{Pas}$ , $\mu_{V} = \SI{9.272e-6}{Pas}$ and \newline
$\mu_{G} = \SI{1.837e-5}{Pas}$\,.

\subsection{Implicit LES }
\label{subsec:implicit_LES}
In this investigation the implicit LES model for cavitating two-phase flows proposed in \citet{Egerer:2016it} is extended for an additional gas phase. The mass fraction of the additional gas phase acts as an active scalar that strongly contributes to the combined equation of state (see previous subsection \ref{subsec:thermo}). In order to avoid unphysical oscillations of the pressure, a thermodynamically consistent coupling of the dependent quantities pressure, density and gas mass fraction has to be ensured. In particular, across simple one-dimensional stationary or moving contact waves in gas mass fraction, pressure and velocity have to remain constant. In order to achieve these properties, the original sensor functional proposed in \citet{Egerer:2016it} requires an extension to detect such contact waves. Furthermore, suitable reconstruction procedures are needed to compute thermodynamically consistent fluxes, especially with respect to the newly added flux for the gas phase.
In the following, a short description of the original scheme and its modifications is presented.
The original ILES flux provides a flux function that operates on reconstructed left- and right-hand primitive variables $\boldsymbol{\Phi}= (\rho, p, \boldsymbol{u}, \xi_{G})$. The smoothness of the flow field is characterized by a sensor functional $f(\vartheta)$ which controls the type of the reconstruction procedure for the specific primitive variable on the cell face $\boldsymbol{\Phi}^{c}$ as follows:
  \begin{equation} 
     \boldsymbol{\Phi}^{c}=(1-f(\vartheta))\boldsymbol{\Phi}^{cc}+f(\vartheta))\boldsymbol{\Phi}^{cu}. 
     \label{eqn:reco} 
  \end{equation}
For smooth flow, linear 4th order reconstruction procedures, indicated with the superscript $cc$, are used for the velocity components and for the pressure field, while the density filed is reconstructed by a linear 2nd order central approximation. In case of non-smooth flow fields, a upwind biased reconstruction is employed indicated with $cu$. There the velocity components are reconstructed using a 3rd order slope limiter \citet{Koren:1993} and the thermodynamic quantities $\rho$, $p$ are reconstructed using the 2nd order minmod slope limiter \citet{Roe:1986}; see \citet{Egerer:2016it}.

Shock and expansion waves are detected by the vorticity-dilation sensor proposed by \citet{Ducros:1999td}:
  \begin{equation}
     \vartheta^{D}=\frac{(\nabla\cdot \boldsymbol{u})^2}{(\nabla\cdot \boldsymbol{u}))^2+(\nabla \times \boldsymbol{u}))^2+\epsilon},
     \label{eqn:sensor1}
  \end{equation}
where $\epsilon$ denotes a very small value to avoid division by zero.

Two-phase pseudo boundaries are detected with the variation of the vapor volume fraction $\gamma = \alpha$ in all three spatial directions here indicated with $i,j,k$:
  \begin{equation}
      \vartheta^{\gamma}=var_{i}(\gamma)+var_{j}(\gamma)+var_{k}(\gamma)\;, 
      \label{eqn:sensor1}
  \end{equation}
where the variation is, e.g. for $i$ evaluated as 
  \begin{equation}
     var_{i}(\gamma)=\| \gamma_{i} -  \gamma_{i-1} \|+\|\gamma_{i+1} -  \gamma_{i} \| . 
     \label{eqn:sensor2}
  \end{equation}
For this investigation the sensor functional was extended for the additional gas phase by adding an additional loop for the 3-D variation in gas volume fraction, $\gamma = \beta_{G}$. If at least one sensor component exceeds its threshold value, the scheme switches to the upwind-biased reconstructions. Thus, the proposed sensor functional is defined as follows:
  \begin{equation}
     f(\vartheta^{D},\vartheta^{\alpha},\vartheta^{\beta_{G}})=
     \left\{
     \begin{array} {rcl}1 , & if & \vartheta^{D}>\vartheta^{D}_{th} ~\| ~ \vartheta^{\alpha} > \vartheta^{\alpha}_{th} ~\| ~  \vartheta^{\beta_{G}}>\vartheta^{\beta_{G}}_{th} \\
     0 , & else. \end{array}\right.
     \label{eqn:sensor3}
  \end{equation}
The threshold values $\vartheta^{D}_{th}=0.95$ and $\vartheta^{\alpha}_{th}=0.25$ are taken from \citet{Egerer:2016it} and the one for the volume fraction of gas is set to $\vartheta^{\beta_{G}}_{th}=0.4$.
 
In addition to the adaptation of the sensor functional, a reconstruction procedure for the gas mass fraction is derived. Instead of using independent reconstruction procedures for density, pressure and gas mass fraction, a consistent coupling is achieved by computing the resulting gas mass fraction using the reconstructed pressure $p^c$ and density $\rho^c$ together with the coupled equation of state $\xi_{G}=EOS(p^c, \rho^c)$. A series of specifically designed test-cases demonstrated that the proposed modifications are sufficient to avoid pressure oscillations while keeping contact waves crisp without artificial smearing.

\nomenclature[G]{$\rho$}{Density\nomunit{[\si{kg/m^3}]}}
\nomenclature[G]{$\alpha$}{Volume fraction vapor \nomunit{[-]}}
\nomenclature[G]{$\beta$}{Volume fraction \nomunit{[-]}}
\nomenclature[G]{$\xi$}{Mass fraction \nomunit{[-]}}
\nomenclature[G]{$\mu$}{Dynamic viscosity \nomunit{[\si{Pa\,s}]}}
\nomenclature[A]{$p$}{Pressure \nomunit{[\si{Pa}]}}
\nomenclature[G]{$\delta$}{Kronecker-Delta \nomunit{[-]}}
\nomenclature[A]{$R$}{Specific gas constant \nomunit{[\si{J/(kg\,K)}]}}
\nomenclature[A]{$\mathrm{Temp}$}{Temperature\nomunit{[\si{K}]}}
\nomenclature[A]{$c$}{Speed of sound \nomunit{[\si{m/s}]}}
\nomenclature[I]{$LV$}{Liquid-Vapor Mixture}
\nomenclature[I]{$G$}{Gas}
\nomenclature[I]{$L$}{Liquid}
\nomenclature[I]{$V$}{Vapor}
\nomenclature[I]{$sat$}{Saturation}
\nomenclature[I]{$th$}{Threshold}
\nomenclature[S]{$cc$}{Central reconstruction}
\nomenclature[S]{$cu$}{Upwind biased reconstruction}
\nomenclature[A]{$\boldsymbol{U}$}{Vector of the conservative variables}
\nomenclature[A]{$\boldsymbol{C}$}{Convective part of the flux vector}
\nomenclature[A]{$\boldsymbol{S}$}{Surface stresses}
\nomenclature[A]{$\boldsymbol{u}$}{Velocity vector field \nomunit{[\si{m/s}]}}
\nomenclature[G]{$\boldsymbol{\tau}$}{Viscous stress tensor \nomunit{[\si{Pa}]}}
\nomenclature[G]{$\vartheta$}{Flow sensor \nomunit{[-]}}
\nomenclature[G]{$\epsilon$}{Very small value \nomunit{[-]}}
\nomenclature[G]{$\boldsymbol{\Phi}$}{Vector of primitive variables}

\section{Numerical setup}
\label{sec:numerical_setup}

Our numerical setup is based on the experiments performed by \citet{Sou:2014hja} and \citet{Bicer:2015cp}. In the experiments tap water is discharged trough a step nozzle into ambient air. Six different operating points were investigated and the cavitation pattern and jet characteristics were visualized with high speed cameras. Additionally, Laser Doppler Velocimetry (LDV) measurements were conducted for one representative operating point. The investigated operating points are characterized by the cavitation number, which is defined here as:
	\begin{equation}
	  \sigma=\frac{p_{out}-p_{sat}}{0.5\,\rho\,\bar{u}^2} \,.
	  \label{eq:sigma}
	\end{equation} 
$p_{out}$ denotes the pressure at the outlet, $p_{sat}$ the saturation pressure of the liquid, $\rho$ the mean density and $\bar{u}$ the mean velocity in stream-wise direction. The tendency for cavitation increases with a lower cavitation number. By changing the flow rate and thus the mean velocity, the cavitation number can be tuned in the experiments. Another important non-dimensional parameter is the Reynolds number 
	\begin{equation}
	  Re=\frac{\bar{u}\,W_{N}}{\nu}, 
	  \label{eq:Re}
	\end{equation}
which relates the inertial forces to the viscous forces. $W_{N}$ stands for the width of the nozzle and $\nu$ denotes the kinematic viscosity. For the investigated operating points, the Reynolds number $Re$ is in the order of 30,000. 

	\begin{figure}[!htb]
		\centering
		\subfigure[]{\includegraphics[height=6.5cm]{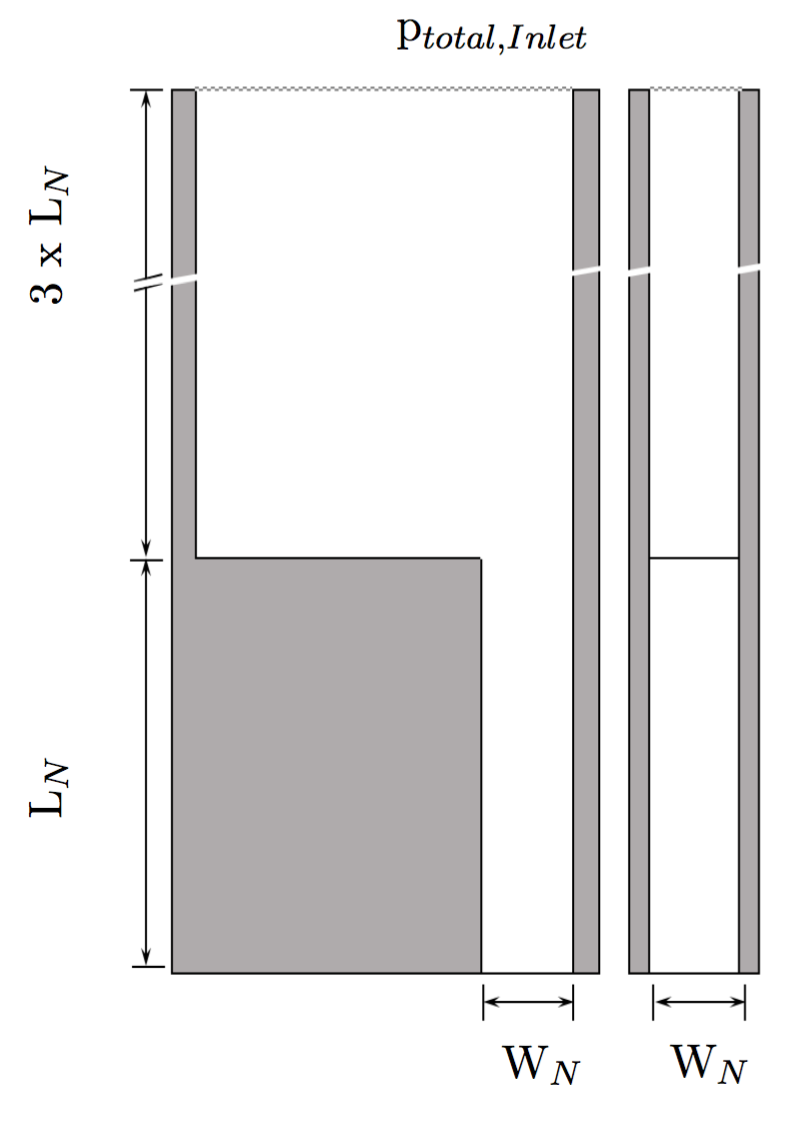}}
		\hspace*{2cm} 
		\subfigure[]{\includegraphics[height=6.5cm]{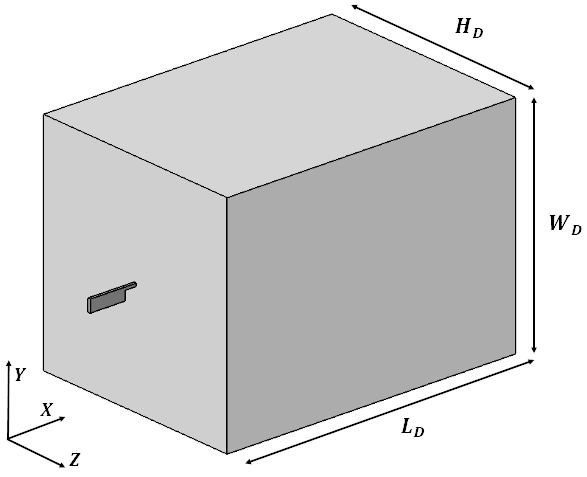}}
		\caption{Sketch of the experimental setup. (a) Upstream region and nozzle in top (left) and side (right) view (b) Full Simulation domain. Figure in (a) is taken from \citet{Trummler:2018vb} Copyright 2018 ASME. }
		\label{fig:setup}
	\end{figure} 

\begin{figure}[!tb]
	 \centering
	 \subfigure[]{\includegraphics[width=0.49\linewidth]{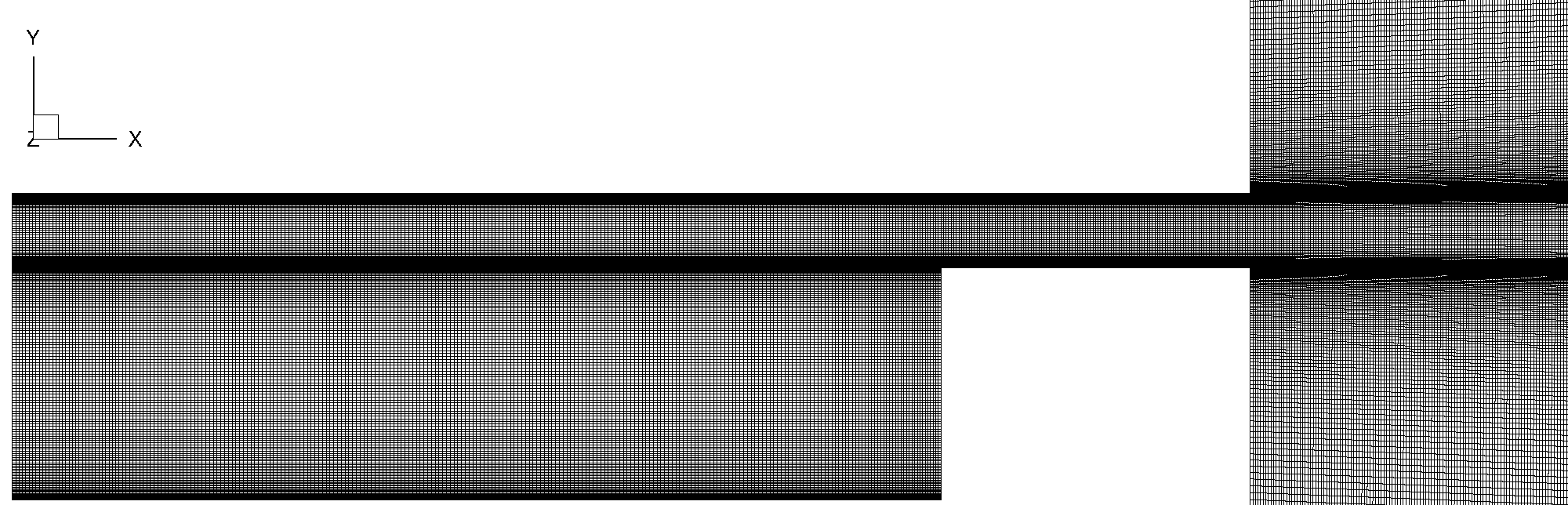}}
	 \subfigure[]{\includegraphics[width=0.49\linewidth]{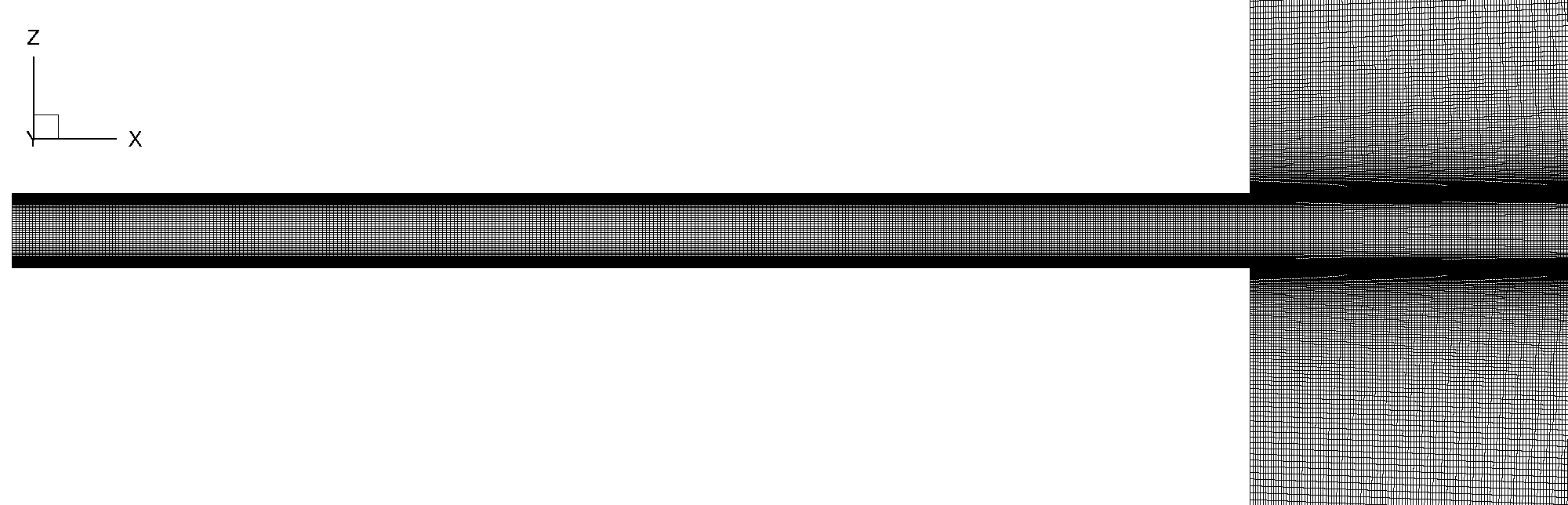}}
	  \caption{Grid on the finest level, every 4th grid line shown. (a) Top view. (b) Side view. }
	 \label{fig:grid}
	\end{figure}

For our numerical studies we selected two representative operating points: $\sigma =1.19$ with \textit{developing cavitation} and $\sigma=0.83$ with \textit{super cavitation}. The numerical setup is depicted in Fig.~\ref{fig:setup}. The geometric measures were taken from the experimental setup: $L_{N}= \SI{8}{mm}$ and $W_{N}= \SI{1.94}{mm}$. The upstream region is elongated to three times the nozzle length, i.e. \SI{24}{mm} and to minimize the influence of boundary conditions on the primary jet break-up, a large outlet domain is added, see Fig.~\ref{fig:setup}~(b). The following factors with respect to the nozzle dimensions are chosen: $L_{D} =25\;\mathrm{x}\;L_{N}$ and $W_{D} =H_{D} =75\;\mathrm{x}\;W_{N}$. We discretize the domain using a block-based, structured mesh. A grid sequencing strategy is applied to reduce the computational cost. This means that for each case we let the flow field first develop on a coarse grid and then refine the grid over several refinement steps to the fine grid. For this simulation five different grid levels were used, where the coarsest grid contains 2.4 million cells and the finest 51.5 million cells. In the finest grid, the smallest cell size in wall-normal direction is \SI{2.5}{\mu m} and the biggest cells in the nozzle are \SI{30.5}{\mu m} long. The grid is also refined around the sharp edge at the nozzle inlet and in the nozzle outflow area where primary jet break-up occurs. Fig.~\ref{fig:grid} depicts slices through the grid, showing every fourth grid line. 

For the inlet boundary condition we use a total pressure condition as also done by \citet{Koukouvinis:2016boa}. The correct pressure for the chosen inlet length is first determined with preliminary simulations and applied later. For $\sigma=1.19$ we prescribe a total pressure of \SI{2.37e5}{Pa} and for $\sigma=0.83$ of \SI{3.03e5}{Pa}. At the outlet we define an outlet pressure of $p_\mathrm{outlet}=\SI{e5}{Pa}$. All walls are treated as viscous isothermal walls. Initially, the whole domain has a pressure of \SI{1e5}{Pa} and a velocity of $u = \SI{0}{m/s}$. Up to the nozzle outlet, the domain is initialized with liquid only $\xi_{G}=0$ and in the outlet region with gas $\xi_{G}=1$.

We set the Courant-Friedrichs-Lewy (CFL) to \num{1.4}, resulting in a timestep of approximately \SI{0.7e-9}{s} on the finest grid. Statistical averaging on the finest grid was performed over \SI{4}{ms}, sampling every time step. The analyzing time of \SI{4}{ms} corresponds to roughly 6 flow through times, depending on the operating point. Mass flow and average velocity at the nozzle inlet and at the outlet were monitored during the simulations. Additionally, the integral vapor content in the nozzle was monitored. 

\begin{figure}[!htb]
	 \centering
	 \subfigure{\includegraphics[width=0.6\linewidth]{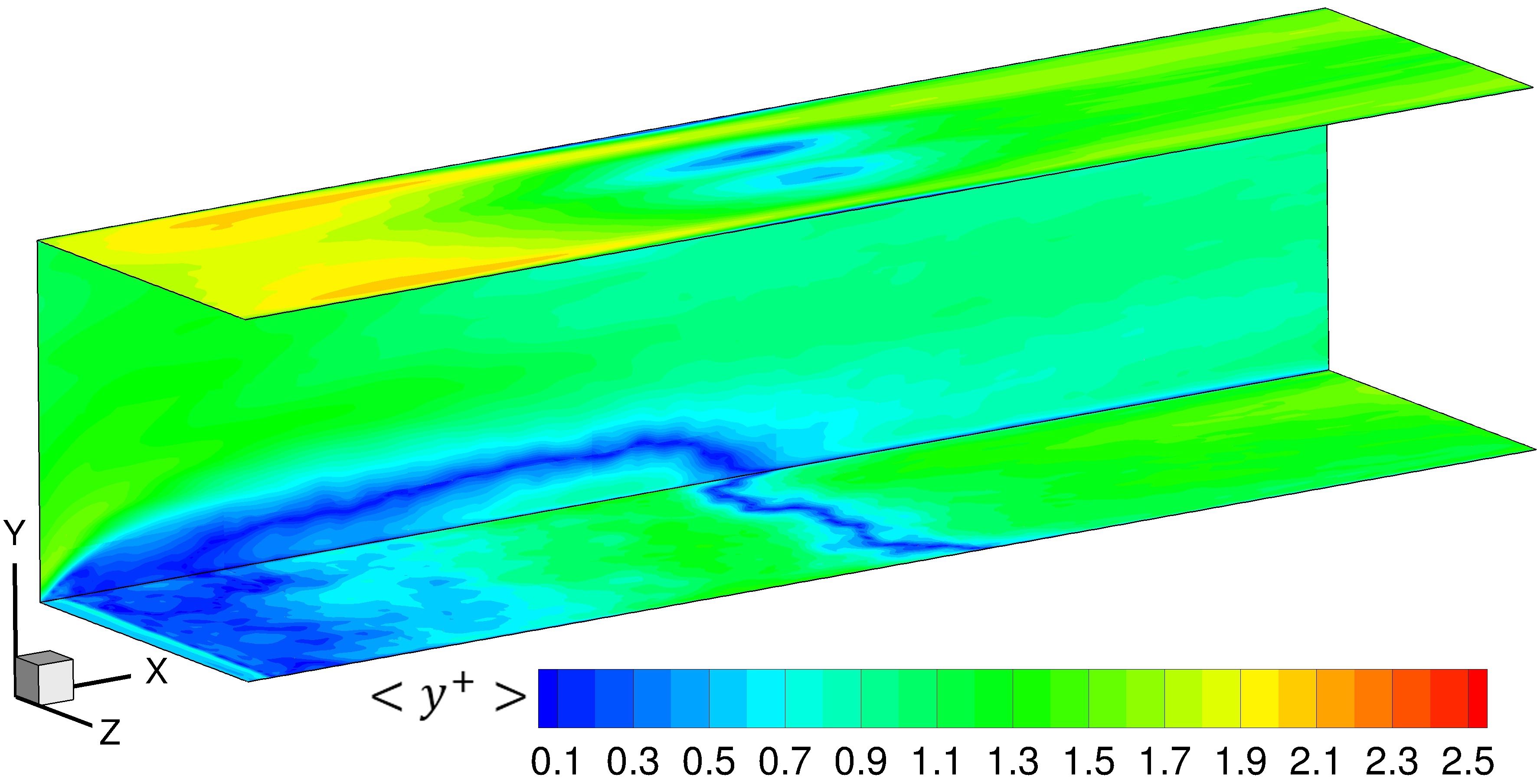}}
	  \caption{Averaged $y^{+}$ for $\sigma=1.19$. (Color online) }
	 \label{fig:y_plus}
	\end{figure}

Fig.~\ref{fig:y_plus} shows the dimensionless wall normal resolution $y^{+}$ for the case $\sigma=1.19$.  $y^{+}$  is the ratio of the wall normal distance of the first grid point to the viscous length scale $\delta_{\nu}$:  
	\begin{equation}
	  y^{+}=\frac{y}{\delta_{\nu}} \; \text{with} \; \delta_{\nu}=\frac{\nu}{u_{\tau}}= \frac{\nu}{\sqrt{\tau_{W} / \rho}} \,. 
	  \label{eq:Re}
	\end{equation}
The near wall resolution in the nozzle is $y=\SI{2.5}{\mu m}$. $\delta_{\nu}$ is calculated using time averaged data, where $\nu$ stands for the average kinematic viscosity and $u_{\tau}$ for the friction velocity. $u_{\tau}$ is the square root of the time averaged ratio of the wall shear stress $\tau_{W}$ over the density $\rho$. As shown in Fig.~\ref{fig:y_plus},  $y^{+}$ in the nozzle is around 1. This indicates that the first grid points lie within the viscous sublayer. Additionally, the chosen grid resolution is motivated by the simulations conducted by \citet{Orley:2015kt} for a similar configuration with slightly higher velocities, where they proved grid convergence for a near wall grid resolution of $\SI{3.9}{\mu m}$. \citet{Koukouvinis:2016boa} demonstrated grid convergence for the same reference case with the identical near wall resolution of $\SI{2.5}{\mu m}$.  

The simulations were conducted with our in-house code CATUM (Cavitation Technical University Munich) developed at the Chair of Aerodynamics and Fluid mechanics, as described in Sec.~\ref{sec:mathe_phys_model}.

\nomenclature[A]{$y^{+}$}{Dimensionless wall normal resolution \nomunit{[-]}}
\nomenclature[G]{$\sigma$}{ Cavitation number \nomunit{[-]}}
\nomenclature[A]{$Re$}{Reynolds number \nomunit{[-]}}
\nomenclature[A]{$u_{\tau}$}{Friction velocity \nomunit{[\si{m/s}]}}
\nomenclature[G]{$\delta_{\nu}$}{Viscous length scale\nomunit{[\si{m}]}}
\nomenclature[G]{$\nu$}{Kinematic viscosity \nomunit{[\si{m^2/s}]} }
\nomenclature[G]{$\tau_{W}$}{Wall shear stress \nomunit{[\si{Pa}]} }

\section{Results }
\label{sec:results}

In the following, the computational results obtained at the finest grid level are presented and discussed. First, the simulation method is validated against experimental data. Then the interaction of cavitation with the mass flow is evaluated and in this context also frequency analysis and Fast Fourier Transformations (FFT) of temporal signals are presented. In subsection \ref{subsec:erosion} the spatial distribution and intensity of collapse events are assessed. Finally, at the end of the section the effects of cavitation and gas entrainment on the liquid jet are investigated. 

\subsection{Comparison with experimental data and validation of the method }
\label{sec:comp_exp}

\begin{figure}[!htb]
\centering
  \centering
  \subfigure[]{\includegraphics[height=7cm ]{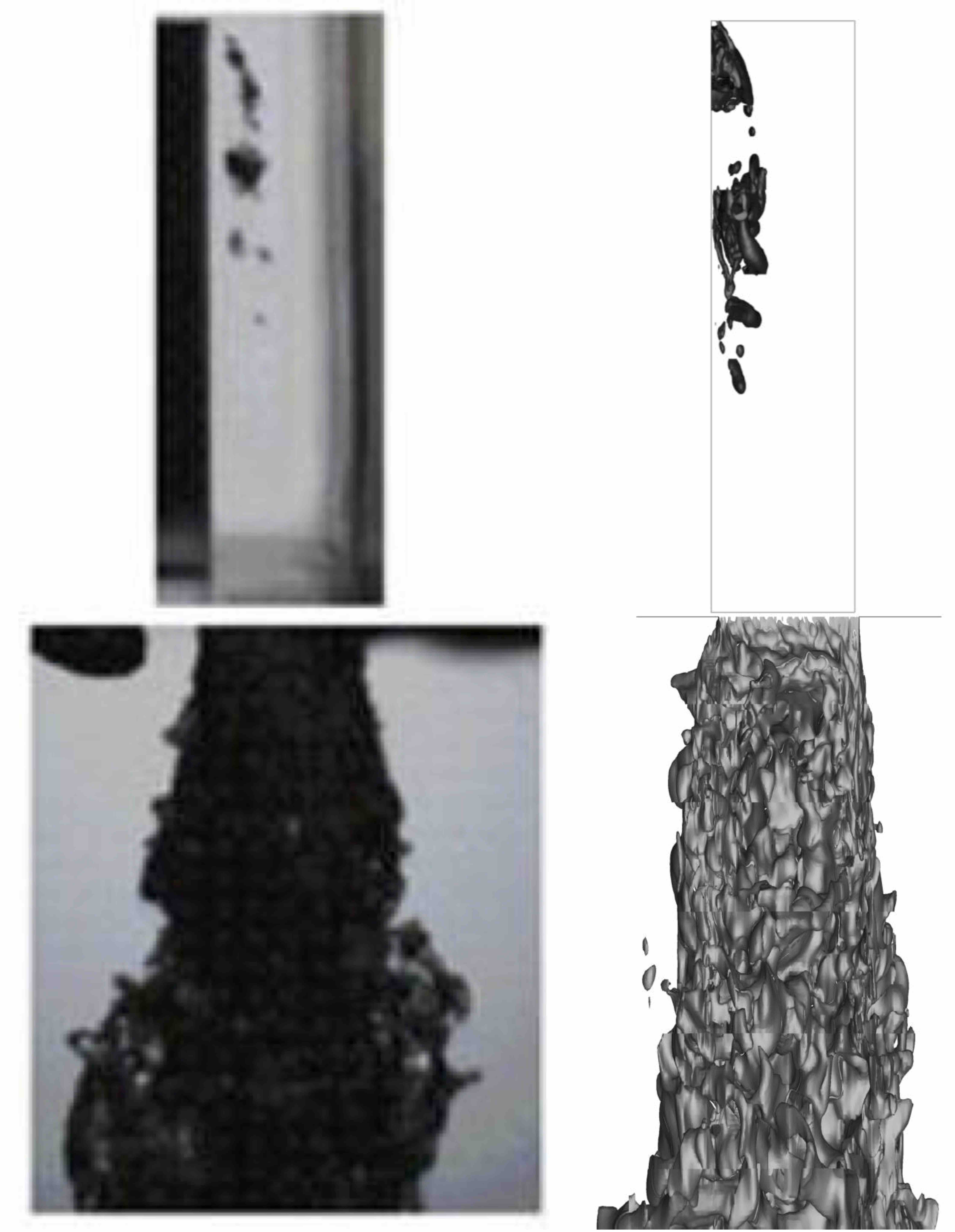}}
  \hspace{2cm}
  \subfigure[]{\includegraphics[height=7cm ]{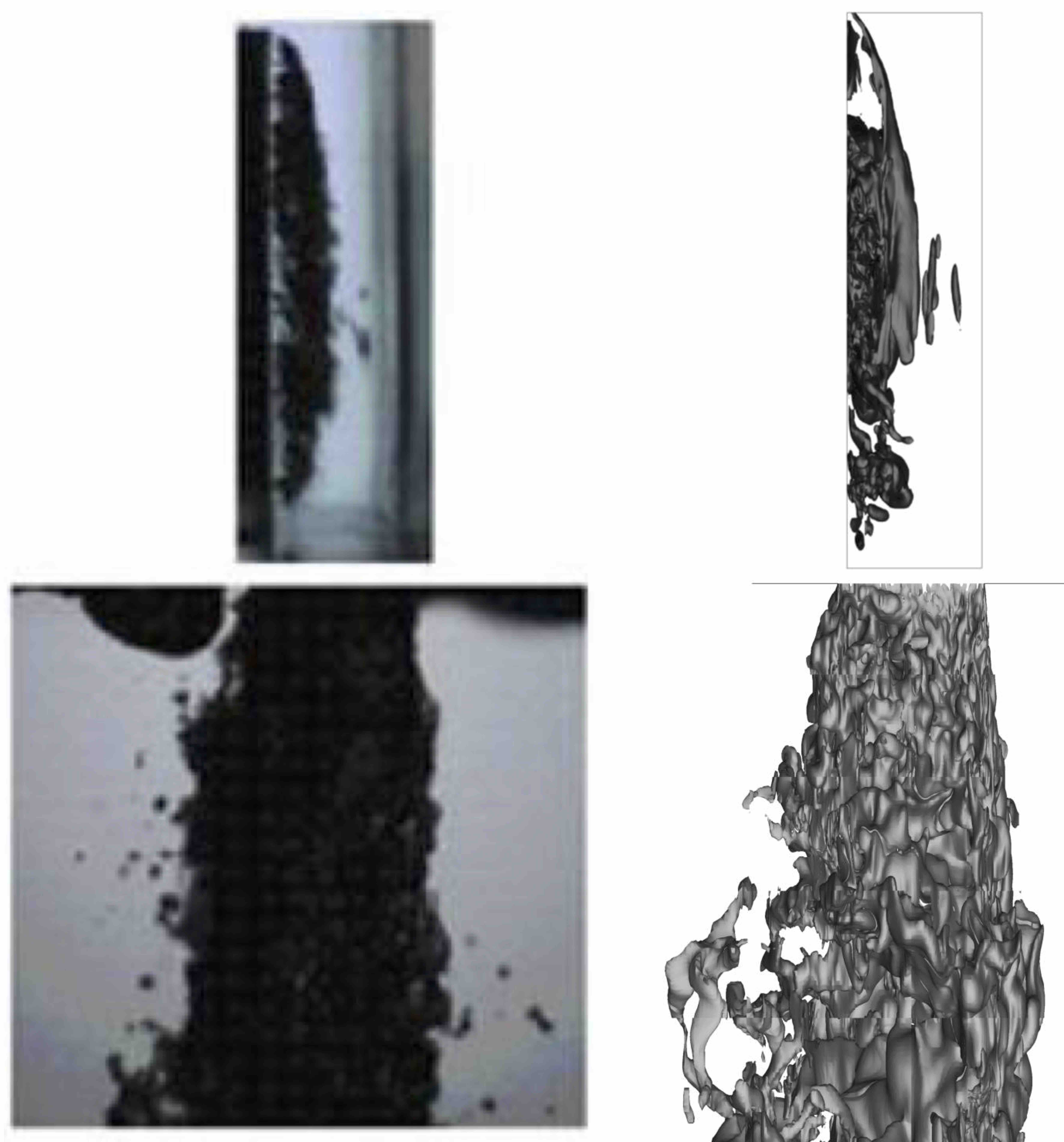}}
  \caption{Comparison of the experimental data from \citet{Sou:2014hja} and the simulation results. Iso-surfaces: Vapor: $\alpha = 0.1$ and jet surface: $\beta_{G}=0.75$. (a) $\;\sigma= 1.19 $ (b) $\;\sigma= 0.84$ ($\sigma_{\mathrm{Experiment}}= 0.83$). Experimental data is reprinted from \citet{Sou:2014hja} with permission from Elsevier (Copyright 2014). Simulation results are adopted from \citet{Trummler:2018vb}, Copyright 2018 ASME. } 
\label{fig:comp_exp_sim}
\end{figure}

In the reference experiment conducted by \citet{Sou:2014hja}, light transmission images of the cavitation in the nozzle and the jet were taken. Fig.~\ref{fig:comp_exp_sim} compares these with our numerical results. In general, there is a good agreement of the cavitation pattern and of the jet characteristics. 

For the higher cavitation number $\sigma= 1.19 $ the experiment predicted \textit{developing cavitation}. In agreement with the experiment, we observe vapor structures in the first part of the nozzle while the jet is not affected by cavitation. 

For $\;\sigma= 0.84$ the cavitation has increased significantly, and the vapor region is spanning from the nozzle inlet up to shortly before the nozzle outlet. In this \textit{super cavitation} regime, the vapor structures collapse close to the nozzle outlet and the experiments showed that an enhanced jet break-up occurs. Our simulation results reproduce this effect. The specification of a constant total pressure at the inlet results at this operating point in a slightly different 
average velocity (\SI{15.2}{m/s} instead of \SI{15.4}{m/s}) and thus in a slightly higher cavitation number.   
\begin{figure}[!htb]
\centering
  \includegraphics[width=0.99\linewidth]{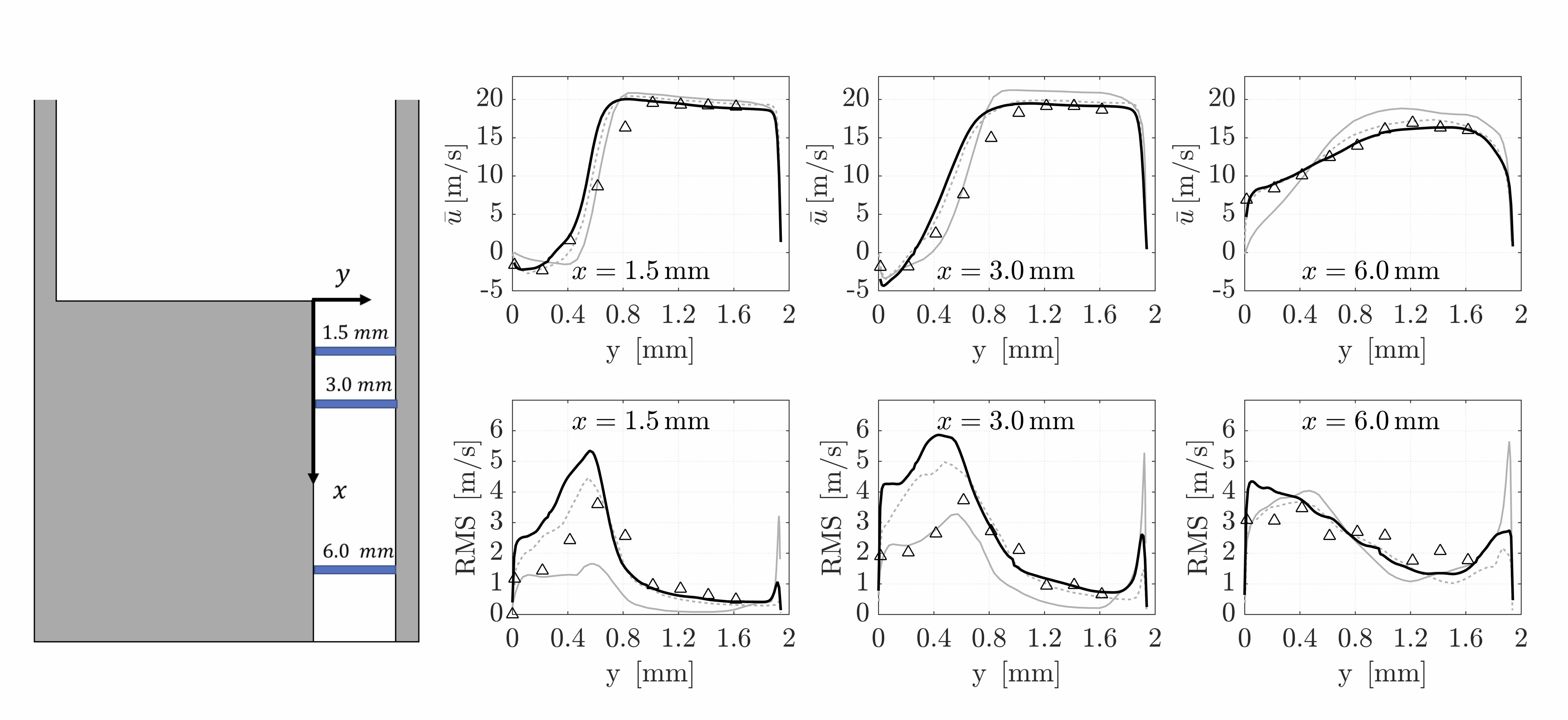}
\caption{Comparison of the experimental data from \citet{Sou:2014hja} (triangles) and the numerical results from our simulations (\textbf{---}); from \citet{Sou:2014hja} ({\textcolor{Gray_Matlab}{---}}) and from \citet{Koukouvinis:2016boa} ({\textcolor{Gray_Matlab} {$-\,-$}}). Top: Mean stream-wise velocity at different positions in the duct. Bottom: RMS (root mean square) of the fluctuations of the velocity in stream-wise direction at different positions in the duct. }
\label{fig:comp_mean_fluc}
\end{figure}

In the experiment \citet{Sou:2014hja} performed LDV measurements for $\sigma=1.19$ and determined the mean stream-wise velocity and the fluctuations in the duct. They evaluated the flow field at three different positions, marked in the sketch in Fig.~\ref{fig:comp_mean_fluc}, in stream-wise direction. In order to validate our numerical method, Fig.~\ref{fig:comp_mean_fluc} compares our results with the experimental data and other numerical results. For comparison, the results from \citet{Sou:2014hja} with the Smagorinsky model and from \citet{Koukouvinis:2016boa} with the wall adapting local eddy viscosity model (WALE) in combination with a barotropic cavitation model are added to the figure. In the upper row of Fig.~\ref{fig:comp_mean_fluc} the mean stream-wise velocity is plotted against the horizontal position. These profiles can be well reproduced with our simulation. Only the detached flow region is slightly wider in the experiment, especially at the position $x= \SI{3}{mm}$. Our results are close to the numerical ones obtained by \citet{Koukouvinis:2016boa}. Additionally, it is worth noting that the average interface velocity can be estimated with the potential flow theory as $ u_{\mathrm{interface}}=u_{\infty}\sqrt{1+\sigma}$, which give a good approximation with \SI{18.94}{m/s}.

The fluctuations in stream wise direction are compared in the bottom row of Fig.~\ref{fig:comp_mean_fluc}. A clear maximum of the fluctuations is visible in the region of the cavity around $y=\SI{0.4}{mm}$ at the positions $x=\num{1.5}$ and $\SI{3}{mm}$. Within the cavitation zone ($y<\SI{0.4}{mm}$) the fluctuations are higher than in the main flow field. Our simulation and the numerical findings by \citet{Koukouvinis:2016boa} overestimate fluctuations compared to the experiment. This may be related to the measurement method, since LDV measurements employ tracing of passive particles. Due to inertia effects the particles mostly remain in the liquid phase, which affects the results of this measurement method. In the main flow field there is an excellent agreement of the fluctuations in our simulation and the experiment. At the right wall (max $y$) the fluctuations increase again.  

Overall there is a good agreement of the simulation results with experimental data, which demonstrates the applicability of our numerical method and thermodynamic modeling for the investigated flows. 

\subsection{Cavitation dynamics }
\label{subsec:cav_dyn}

\begin{figure}[!tb]
  \centering
  \includegraphics[width=0.99\linewidth]{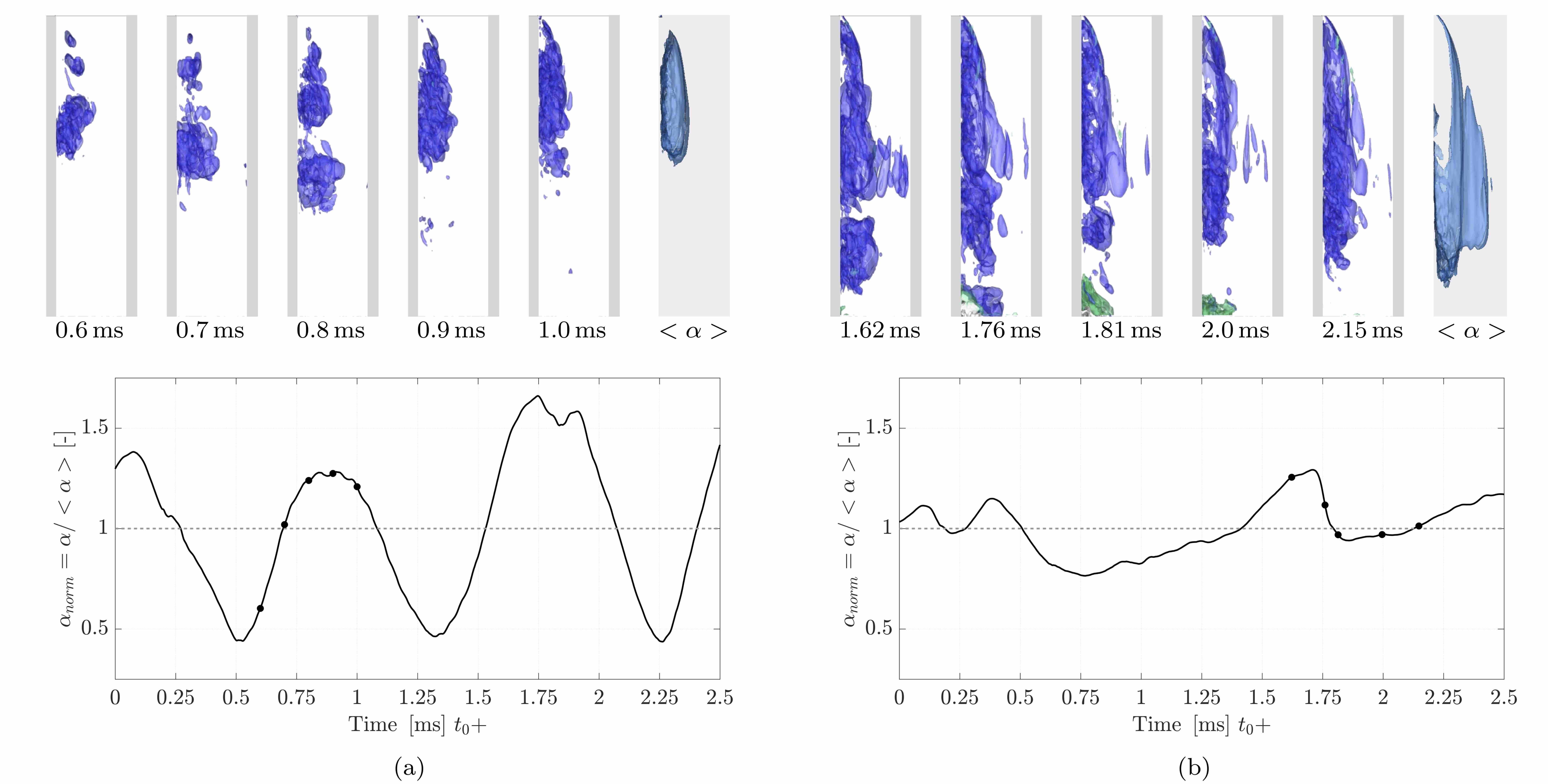}
 \caption{Top: time series for $\;\sigma= 1.19 $  (a)  and for $\;\sigma= 0.84 $ (b). Vapor in the nozzle (blue, 10$\%$) and gas sucked into the nozzle (green, 10$\%$). The last pictures show the temporal averaged vapor content. Bottom: corresponding signals of the integral normalized vapor content in the nozzle. Depicted time steps are marked by dots. (Color online)}
  \label{fig:cav_dyn_jet}
\end{figure}

In order to analyze cavitation dynamics, time series for $\sigma = 1.19$ and $\sigma = 0.84$ are shown in Fig.~\ref{fig:cav_dyn_jet}. At $\sigma = 1.19$ (Fig.~\ref{fig:cav_dyn_jet}~(a)) cavitating span-wise vortices form at the shear layer at the nozzle inlet, as depicted at $t=\SI{0.6}{ms}$. Then these span wise vapor structures coalesce to bigger ones and start to form a vapor sheet. At the end of the sheet, vapor clouds detach and are advected downstream where they collapse; see $t=\SI{0.8}{ms}$ and $t=\SI{0.9}{ms}$. 
A clear correlation between the integral vapor volume (plotted in the bottom row) with the instantaneous snapshots is  visible. For $\sigma=1.19$ we observe a periodic oscillation with a relatively high normalized amplitude of approximately $50\%$ compared to the averaged vapor content of 2.51$\%$. The depicted time instants show the growth phase of the cavity with increasing vapor volume. After the collapse of the detached vapor structures the vapor content diminishes. Spectral analysis of the integral vapor content revealed that the shedding frequency for $\sigma=1.19$ is $f=1110\,\si{Hz}$, which corresponds to a period of $T=0.9\,\si{ms}$, see subsection~\ref{subsec:frequ}. 

Fig.~\ref{fig:cav_dyn_jet} also includes graphics with the 10$\%$ iso-surface of the time averaged vapor content in the nozzle. We observe that for $\sigma = 1.19$ in average no vapor is present directly at the nozzle inlet edge  beyond the separated flow. The average cavity length is $l_{cav}=4.0\;\si{mm}$ and the length of the  attached cavity is $l_{acav}=3.3\;\si{mm}$. For comparison, the nozzle length is $8\;\si{mm}$.

For $\sigma = 0.84$\, (Fig.~\ref{fig:cav_dyn_jet}~(b)) there is a stable cavitation sheet visible, at the end of which vapor clouds detach. Additionally, sometimes thin cavitating vortices in stream-wise direction occur. These are also observed in experiments e.g. \citet{mauger2012shadowgraph}. \citet{Egerer:2014wu} discusses the formation of cavitating vortices in such a configuration in detail. 
In the first time step ($t=1.62\;\si{ms}$), a larger vapor structure has detached from the main sheet. Mostly, these vapor structures collapse within the nozzle, but a few large structures also reach the nozzle outlet and subsequently lead to partial entrainment of gas. When a vapor cloud reaches the nozzle outlet, there is a pressure gradient from the outflow region to the vapor region at saturation pressure and gas is sucked into the nozzle. When a detached vapor cloud reaches the outlet, the gas from the outflow area partly entrains the nozzle filling up the cavity of the vapor and  is then stopped by the liquid. Conversely, when a continuous vapor sheet reaches the outlet, gas fills up the entire nozzle length, leading to a complete detachment of the flow and to the so-called 'hydraulic flip'. At the investigated operating point only partial gas entrainment takes place. At $t=1.76\;\si{ms}$ the detached vapor cloud reaches the outlet, which leads to partial gas entrainment from the outflow region into the nozzle ($t=1.81\;\si{ms}$ green isosurfaces).  Afterwards, this gas is pushed out again, and in the last time step no gas is present in the nozzle. 

The time averaged vapor sheet spans over nearly the whole nozzle length and also fills up the area beyond the  separated flow at the nozzle inlet. In this case $l_{cav}=7.2\;\si{mm}$ and $l_{acav}=6.4\;\si{mm}$.  As expected, the dominant frequency decreases for the stronger cavitating case and is at $f = 750\;\si{Hz}$, $T=1.32\;\si{ms}$, see also subsection~\ref{subsec:frequ}. The normalized amplitude of the integral vapor content with approximately 15$\%$ is significantly smaller compared to $\sigma = 1.19$. However, the averaged vapor content nearly quadruples in comparison to $\sigma = 1.19$  with 8.24$\%$. 
 
An important dimensionless quantity for oscillating phenomena is the Strouhal number $St$.  We estimate $St$ using the time averaged length of the attached cavity $l_{acav}$, and the mean stream-wise velocity $\bar{u}$ as 
\begin{equation}
   St =\frac{f\cdot l_{acav}}{\bar{u}}, 
  \label{eq:st}
\end{equation} 
and obtain $St =0.29$ for $\sigma = 1.19$ and $St= 0.32$ for $\sigma = 0.84$, see Table~\ref{tab:overview}. Experiments with cylindrical orifices revealed Strouhal numbers in the  range of $0.3 - 0.5$ \citep{Stanley:2011gr, Sugimoto:2009, Sato:2002vv}. Consequently, our results compare well to  experimental studies for similar configurations. 
\begin{table}[!htb]
\caption{Characterizing parameters of the investigated cavitation regimes.}
\centering\begin{tabular}{|rrrrrrrl|}
\hline
    $\boldsymbol{\sigma}$        \textbf{[-]} & 
    $\boldsymbol{\bar{u}}$       \textbf{[m/s]}  & 
    $\boldsymbol{\bar{\dot{m}}}$ \textbf{[g/s]} & 
    $\boldsymbol{l_{cav}}$           \textbf{[mm]}   & 
    $\boldsymbol{l_{acav}}$        \textbf{[mm]}   & 
    $\boldsymbol{f}$             \textbf{[Hz]}   & 
    $\boldsymbol{St_{\;l_{acav}}}$  \textbf{[-]}   &
     \textbf{Cavitation regime} \\
    \hline
    1.19  & 12.8  & 48.37 &   4.0    & 3.3  & 1110   & 0.29   & developing \\
    0.84  & 15.2  & 57.08 &   7.2    & 6.4  & 750  & 0.32  & super \\
    \hline
    \end{tabular}%
  \label{tab:overview}%
\end{table}%

\nomenclature[A]{$St$}{Strouhal number \nomunit{[-]}}
\nomenclature[A]{$f$}{Frequency \nomunit{[\si{1/s}]}}
\nomenclature[A]{$t$}{Time \nomunit{[\si{s}]}}
\nomenclature[A]{$T$}{Period \nomunit{[\si{s}]}}

\subsection{Effects of cavitation and partial gas entrainment on the mass flow}
\label{subsec:frequ}

In this section the effects of cavitation and partial gas entrainment on the mass flow at the nozzle inlet and outlet are studied. First we evaluate the discharge coefficients and compare them with the experimental ones. Then we analyze the direct interaction of the cavitation on the temporal mass flows and evaluate the dominant frequencies with FFT analysis. Finally, we also investigate the amplitude of the fluctuations of the mass flows. 

The discharge coefficient is defined as 
  \begin{equation}
    C_D= \frac{\dot{m}}{\dot{m}_{\mathrm{theoretical}}}= \frac{\dot{m}}{ A_{\mathrm{theoretical}} \sqrt{2 \rho (p_{\mathrm{inlet}} - p_{\mathrm{outlet}})}}, 
    \label{eq:v_interface}
  \end{equation} 
where $A_{\mathrm{theoretical}}$ denotes the cross section area of the nozzle. It is well known, that cavitation leads to a decrease of the discharge coefficient, see e.g. the experimental investigations by \citep{nurick1976orifice, payri2004influence}. From the available experimental data by \citet{Sou:2014hja}, we can derive the experimental discharge coefficient $C_{D, exp}$ and obtain $C_{D,exp}=0.83$ for $\sigma = 1.19$ and $C_{D,exp}=0.81$ for $\sigma = 0.84$. In our simulations we find $C_{D}=0.83$ for $\sigma = 1.19$, which is in excellent agreement with the experimental data. For $\sigma = 0.84$ our numerical $C_{D}$ is with $C_{D}=0.80$ slightly smaller than the experimental one. 

\begin{figure}[!htb]
\centering
  \subfigure{\includegraphics[width=0.99\linewidth ]{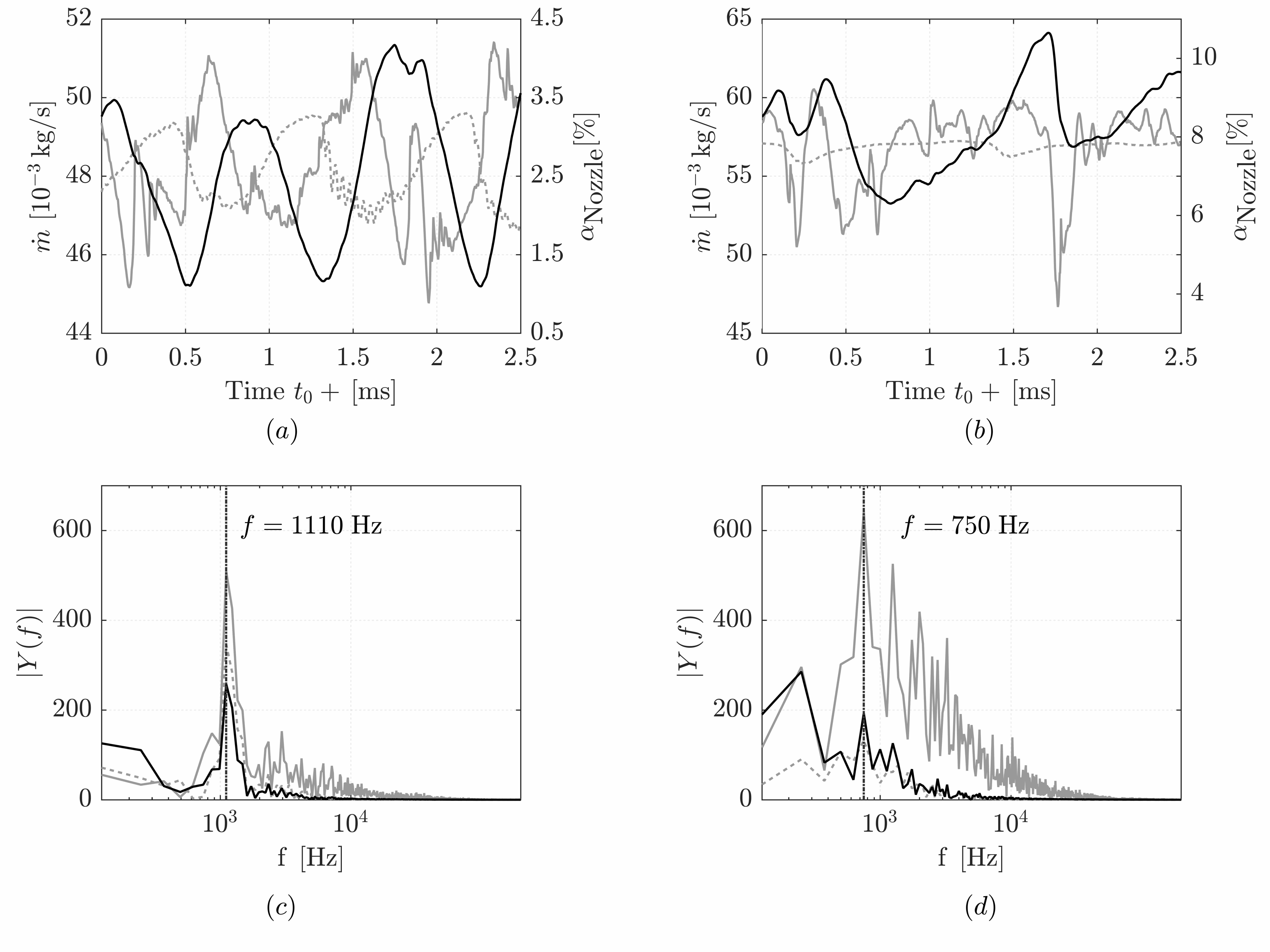}}
   \caption{Temporal evolution of the mass flows $\dot{m}$ at nozzle inlet ({\textcolor{Gray_Matlab}{- - -}}) and outlet ({\textcolor{Gray_Matlab}{---}}) and vapor content $\alpha_{\mathrm{Nozzle}}=\mathrm{V_{Vapor}}/\mathrm{V_{Nozzle}}\cdot100\%$  ({\textcolor{black}{---}}) as well as the amplitude of their complex frequency spectra $|Y(f)|$ over the frequency $f$. The operating point $\sigma = 1.19$ is shown in (a) and (c) and $\sigma = 0.84$ is shown in (b) and (d). }
\label{fig:temp_and_fft}
\end{figure}

We now analyze the temporal signals and investigate the direct effects of the cavitation on the mass flows. Fig.~\ref{fig:temp_and_fft} shows the temporal evolution of the mass flow and the integral vapor content together with their frequency spectra. For both operating points a very strong correlation between the temporal evolution of the three signals is apparent. Additionally, the FFT reveals that the dominant frequency of the mass flows and the vapor volume is identical.  

For $\sigma = 1.19$ we observe a periodic oscillation of the vapor content (black line) due to the shedding, see Fig.~\ref{fig:temp_and_fft}~(a). The mass flow at the inlet (gray dotted line) oscillates  asynchronously to the vapor content. A mass flow maximum at the nozzle inlet corresponds to a rising flank of the vapor content. The vapor formation in the nozzle and the mass flow at the nozzle inlet are strongly coupled. When the mass flow and thus the velocity increases, the static pressure drops locally, and consequently more vapor is generated. The formed vapor  reduces the effective cross-section, which results in a drop of the mass flow at the nozzle inlet. A reduced mass flow leads to less cavitation and less blocking of the cross-section which then consequently allows an increase of the mass flow again. The mass flow at the nozzle outlet (gray line) increases time delayed to the vapor content. The delay is approximately $ 0.6\;\si{ms}$, which is slightly less than the average flow through time of  $ 0.625\;\si{ms}$ ($ 8\;\si{mm}$ nozzle length and average velocity of $12.8\;\si{m/s}$). The formation of vapor leads to a replacement of the liquid and consequently also to a short acceleration of the fluid towards the outlet. This can cause the time delayed peak of the mass flow at the outlet. 

For the super-cavitating case $\sigma=0.84$ (Fig.~\ref{fig:temp_and_fft}~(b)), the quantities oscillate with a lower frequency than $\sigma=1.19$. In the depicted timespan we observe one 'double peak' in the integral vapor content  at the beginning and one normal peak at approximately $t_0\;+\;1.6\;\si{ms}$. Both peaks correspond to a clearly visible, slightly time advanced, drop in the mass flow at the nozzle inlet. Thus, as for  $\sigma=1.19$, the increase of the vapor content initiates a drop of the mass flow at the inlet. For $\sigma=0.84$, the mass flow at the outlet appears to be stronger governed by the cavitation. Every maximum in the vapor content corresponds to a slightly time delayed minimum in the mass flow at the outlet. At $t_0\;+\;1.6\;\si{ms}$ there is a high peak in the vapor content  and afterwards a significant drop of the mass flow at the outlet. From analysis of the flow field data we find that the high vapor content leads to partial gas entrainment which evokes a back flow and causes the drop of the mass flow. The depicted timespan in Fig.~\ref{fig:temp_and_fft}~(b) is identical with the one in Fig.~~\ref{fig:cav_dyn_jet} (b), and the drop of the mass flow corresponds to the time step with the gas entrainment. 

The amplitude of the spectra is depicted in Fig.~\ref{fig:temp_and_fft}~(c) and (d). For $\sigma = 1.19$, the frequency with the highest amplitude is $f = 1110\;\si{Hz}$, which is equivalent to a period of $T=0.9\;\si{ms}$. The  dominant frequency decreases with the cavitation number and for $\sigma = 0.84$ the peak is at $f = 750\;\si{Hz}$, $T=1.32\;\si{ms}$.  Thus in both cases the dominant frequencies of the mass flows and the vapor content  coincide. This was also observed experimentally by \citet{Duke:2015cta}, who found corresponding peaks of the fractional power spectral density for the fluctuations at the inlet, for void-filled bubbles, the wall film and the cavitation interface. Recently, \citet{Beban:2017vo} demonstrated a correlation between the mass flow at nozzle outlet and integrated vapor content, for a submerged flow. 

The coincidence of the peak frequencies of different quantities can be very useful to estimate unknown quantities. For instance, given an experimental time-resolved mass flow signal, one can evaluate the dominant frequency of the mass flow signal and, one may assume that this frequency is equivalent to the shedding frequency as well. Care has to be taken as spurious frequencies may be generated by the experimental apparatus. Furthermore, with an approximate Strouhal number the cavity length can be estimated. Conversely, based on the cavity length, fluctuations of the mass flow can be estimated with the Strouhal number. 

For most applications, such as combustion processes or injection processes, a stable mass flow of the fluid is desirable. Consequently, fluctuations of the discharged mass should be reduced. In Fig.~\ref{fig:temp_and_fft}~(a) and (b) we see significant fluctuations of the mass flow. Table~\ref{tab:mass_fluc} contains the relevant mass flow data as found in our simulations. Besides the mean mass flow $\overline{\dot{m}}$, the mean deviation of the mass flow from the average at the inlet and at the outlet $\overline{|\dot{m}'|}$ as well as the maximum deviations from the mean values $\mathrm{max}(\dot{m}')$ are evaluated. As indicated in Table~\ref{tab:mass_fluc}, the fluctuations at the inlet are very small on average. The average deviation and the maximum deviation at the inlet are higher for $\sigma = 1.19$. On one hand this is partly due to the lower absolute value of the mean mass flow, but on the other hand the mass flow at the inlet seems to be more strongly affected by cavitation at $\sigma = 1.19$ since the cavitation process occurs closer to the nozzle inlet. However, oscillations at the outlet are mostly more crucial for stable processes. Here we find average deviations approximately twice as large as at the inlet. And the maximum deviations have increased as well, rising with a decreasing cavitation number. As already discussed, the immense instantaneous  drop of the mass flow at the outlet in the case of $\sigma = 0.84$ by nearly $20\,\%$ is triggered by gas entrainment. 
\begin{table}[!htb]
  \centering
  \caption{Mass flows and their fluctuations. }
    \begin{tabular}{|rrrrrr|}
    \hline
     $\boldsymbol{\sigma}$                    \textbf{[-]} & 
     $\boldsymbol{\overline{\dot{m}}}$        \textbf{[g/s]} & 
     $\boldsymbol{\overline{|\dot{m}'_{in}|}}$      \textbf{[\%]} & 
     $\boldsymbol{\mathrm{max}( \dot{m}_{in}' )}$ \textbf{[\%]} & 
     $\boldsymbol{\overline{|\dot{m}'_{out}|}}$     \textbf{[\%]} &  
     $\boldsymbol{ \mathrm{max}( \dot{m}_{out}' )}$\textbf{[\%]}  \\
    \hline
    1.19  & 48.4 & 1.63  & 3.32  & 2.75  & 7.12   \\
    0.84  & 57.0 & 0.42  & 1.93  & 3.39  & 18.12  \\
    \hline
    \end{tabular}%
  \label{tab:mass_fluc}%
\end{table}%

\nomenclature[A]{$C_D$}{ Discharge coefficient \nomunit{[-]}}
\nomenclature[A]{$\dot{m}$}{Mass flow \nomunit{[\si{kg/s}]}}
\nomenclature[A]{$A$}{Area \nomunit{[\si{m^2}]}}
\nomenclature[S]{$'$}{fluctuations}
\nomenclature[A]{$Y$}{Complex frequency spectra \nomunit{[-]}}

\subsection{Flow dynamics associated with collapsing vapor patterns}
\label{subsec:erosion}

For partial cavitation, vapor structures detach and are advected downstream, where they implode due to the increased surrounding pressure. The collapse of a vapor structure leads to a collision of liquid fronts and subsequently to the emission of shock waves. In applications with higher inlet pressures, such as with Diesel injector components, such collapses can lead to severe material damage and to device failure. Available experimental data for this investigation do not provide information about collapse dynamics, collapse locations and possible impact loads. However, our compressible approach enables us to capture shock waves following collapse events. To detect the collapse events, we utilize the collapse detection algorithm developed by \citet{mihatsch2015cavitation}. Collapse events are identified based on the vanishing of the vapor content in the cell and all next-neighbor cells and a negative divergence of the velocity. The maximum pressure is recorded and stored as collapse pressure $p_{c}$. \citet{schmidt2014assessment} and \citet{mihatsch2015cavitation} have shown that using an inviscid flow model the recorded collapse pressure $p_{c}$  depends on the grid resolution, since the collapse pressures are inversely proportional to the cell size at the collapse center. They proposed to scale the pressure with 
	\begin{equation}
	  p_{\mathrm{c, scaled}}=p_c \cdot \frac{V_{\mathrm{cell}}^{1/3}}{l_{\mathrm{ref}}}
	  \label{eq:pwall}
	\end{equation}
where $V_{\mathrm{cell}}$ is the cell volume and $l_{\mathrm{ref}}$ a reference length. The scaling law is motivated by the ideally radial decay of the maximum pressure with $1/r$ of an emitted shock wave. In order to predict collapse pressures correctly, the reference length $l_{\mathrm{ref}}$  needs to be calibrated with experimental data, see \citet{mihatsch2015cavitation}. Since there is insufficient information for calibrating the reference length for this investigation, we have adopted the smallest length scale in the evaluated dataset $l_{\mathrm{ref}}$=7.14\,\si{\mu m}. Consequently, the scaled collapse pressures $p_{\mathrm{c, scaled}}$ do not provide absolute values for the case and can not be compared to material parameters. Nevertheless, the scaling is necessary for grids with inhomogeneous resolution. 

\begin{figure}[!htb]
\centering
 \includegraphics[width=0.99\linewidth]{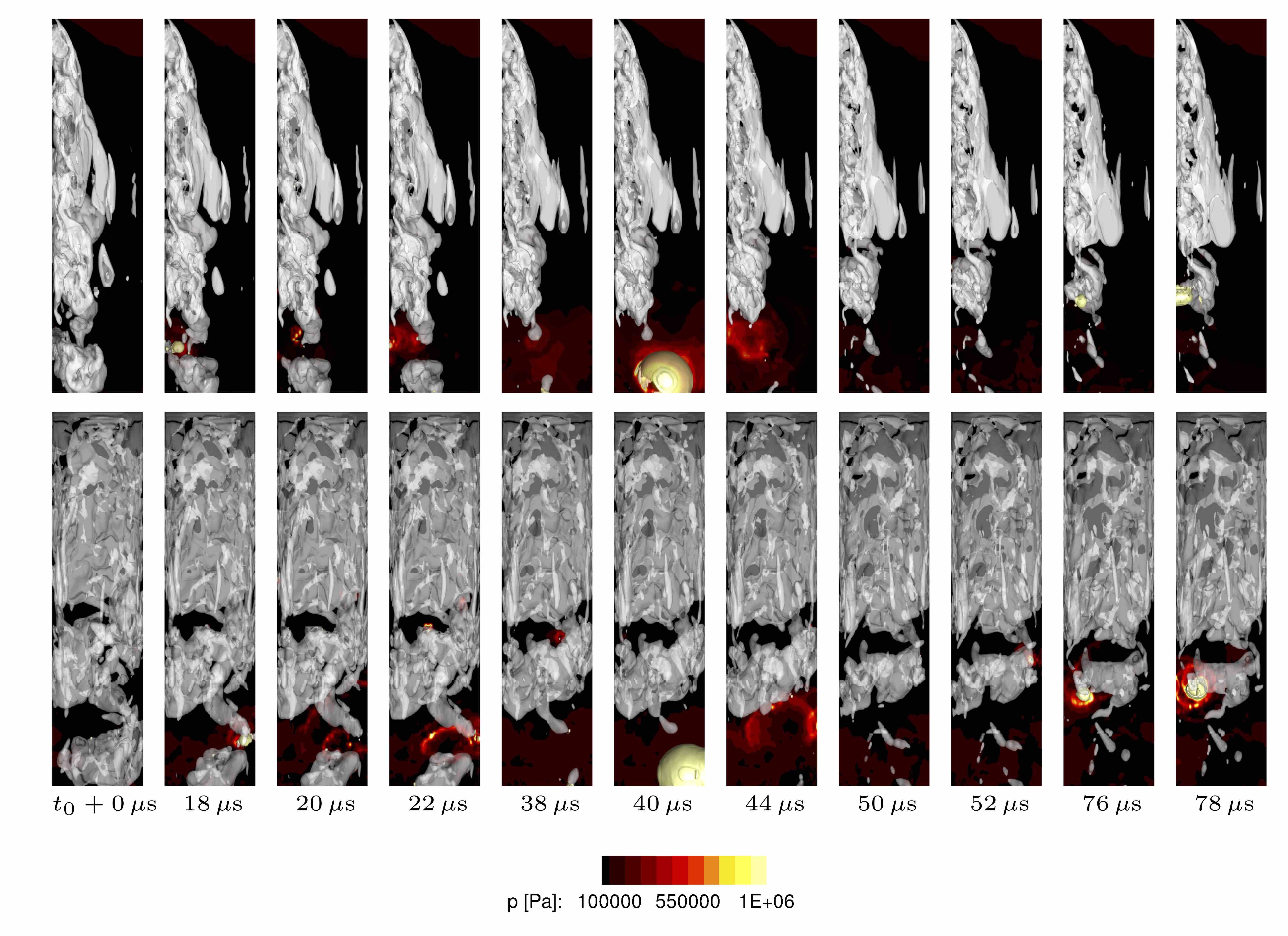}
\caption{Time series for $\sigma= 0.84 $ with  selected time steps capturing collapse events. Iso-surfaces: vapor (10 \%, white) and pressure ($1\cdot10^{6}\; \si{Pa}$, yellow) and wall pressure. Top row: Top view with the wall pressure at the back wall. Bottom row: Side view with wall pressure on the side wall. (Color online)}
\label{fig:timeserie_pc_all}
\end{figure}%

In Fig.~\ref{fig:timeserie_pc_all}, a time series capturing collapse events at $\sigma=0.84$ is depicted. At the end of the vapor sheet and in the detached cloud the vapor structures fragment and eventually collapse. After the collapse they emit a spherical shock wave, resulting in a circular pressure foot print. In the time series, the legend for the pressure is chosen for a lower range to capture and visualize collapse events. The pressure peaks after a collapse event have a very short duration and decay radially with $1/r$, such that it is very difficult to capture a collapse event and its aftermath with regular output. 
\begin{figure}[!htb]
  \centering
   \subfigure{\includegraphics[height=8cm]{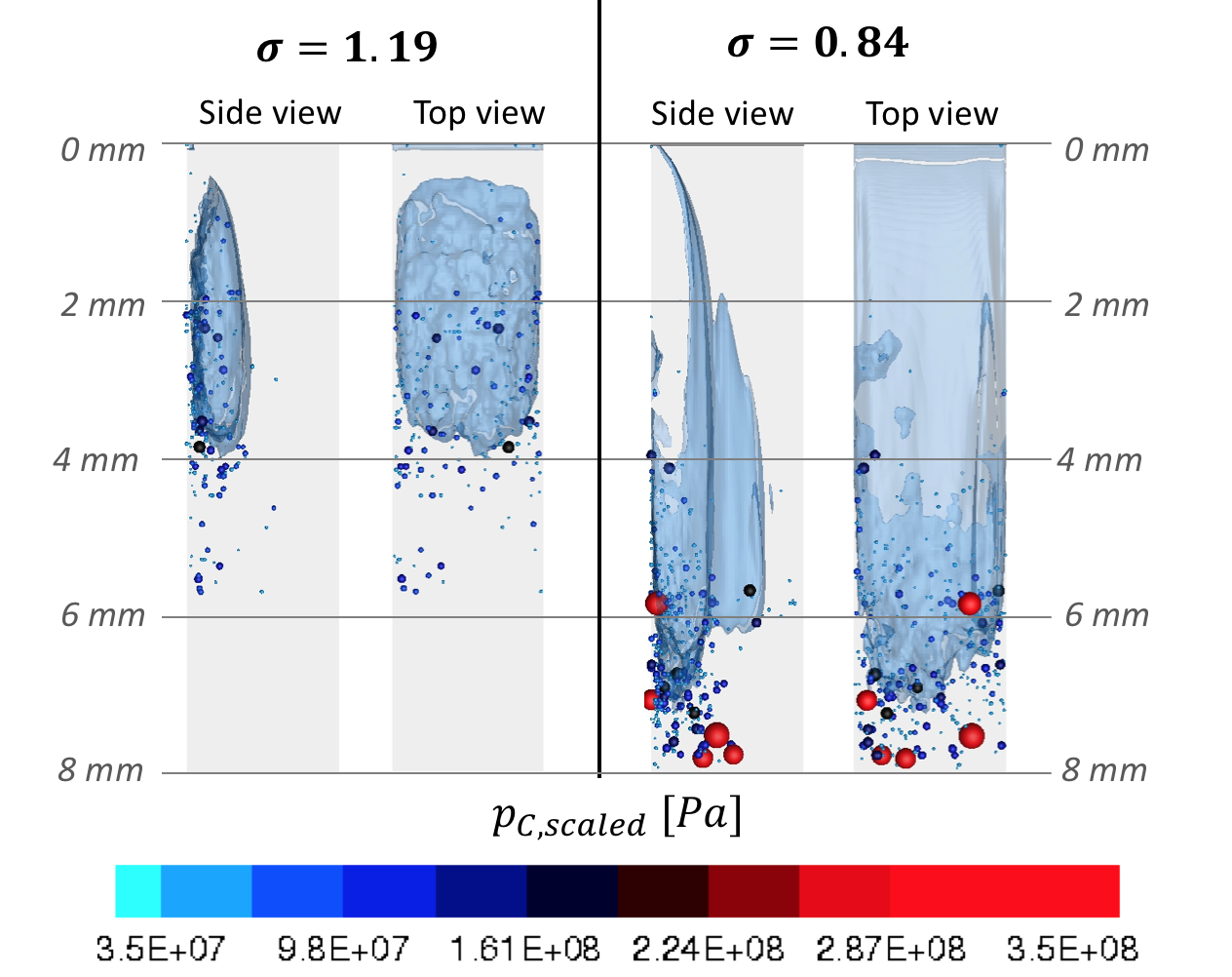}}
 \caption{ Detected collapse events shown as spheres with color and size corresponding to the collapse pressure and iso-surface of the averaged vapor content (10 \%, blue ) over the analysis time of $4\;\si{ms}$. (Color online)}
\label{fig:co_sou3_sou5}
\end{figure}

Fig.~\ref{fig:co_sou3_sou5} shows the detected collapses over the analysis time of $4\;\si{ms}$ together with the averaged vapor content. For$\;\sigma = 1.19$ collapses are recorded between $25\,\%$ and $60\,\%$ of the nozzle length. Mainly they occur close to the wall in the second half of the average cavity, where during the cavitation process the highest fluctuation of the vapor takes place.  At the operating point $\sigma = 0.84$ more collapses are detected and the recorded pressures are significantly higher than $\sigma = 1.19$.  In this case, the collapses mostly occur at the end of the average cavitation zone or even shortly after and thus very close to the nozzle outlet. 
In order to evaluate the validity of the obtained results for high pressure applications with a Diesel fuel, we compare our results to those obtained by \citet{Egerer:2014wu}. They perform LES of turbulent cavitating flows in a micro channel with a typical Diesel-like test fluid and outlet pressures of $O(100\,\si{bar})$. For \textit{developing cavitation}, with high vapor dynamics in the first half of the channel, they monitored collapses in the first half of the channel and close to the wall. For strong cavitation, collapses were mainly detected close to the outlet and often significantly dislocated from the nozzle walls. This comparison indicates that areas where collapse events mainly occur can be estimated based on cavitation regimes independently of the working fluid and the injection pressure. However, in contrast to our results, they have predicted a higher risk of cavitation erosion for the operating point with \textit{developing cavitation}. In our simulations  higher collapse pressures are detected for the smaller cavitation number in the \textit{super-cavitating} regime.

An important aspect is that collapses amplify turbulent fluctuations, as has been observed experimentally \citep{Sou:2007jd} and numerically \citep{Orley:2015kt}. Consequently intense collapses close to the nozzle outlet are considered as one of the main mechanisms to enhance primary jet break up. The effect of cavitation on the primary jet break up is discussed in section \ref{subsec:jet}. 

\nomenclature[I]{$c$}{Collapse}
\nomenclature[A]{$V_{cell}$}{Cell volume \nomunit{[\si{m^3}]} }

\subsection{Effects of cavitation and partial gas entrainment on the jet  }
\label{subsec:jet}

\begin{figure}[!htb]
  \centering
     \includegraphics[width=0.99\linewidth]{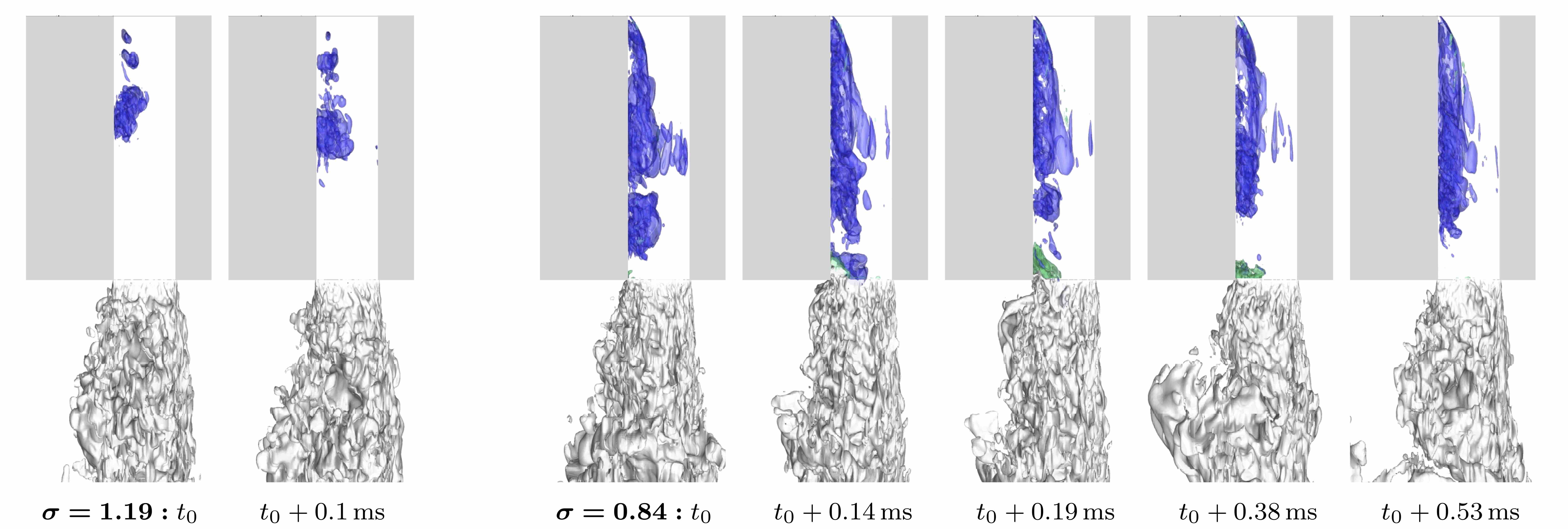}
    \caption{Time series for $\;\sigma= 1.19 $  (left)  and for $\;\sigma= 0.84 $ (right). Vapor in the nozzle (blue, 10$\%$), gas sucked into the nozzle (green, 10$\%$) and jet surface (gray). (Color online) }
  \label{fig:cav_dyn_jet2}
\end{figure}

Intense cavitation in the nozzle affects the flow field close to the nozzle outlet and subsequently the discharged liquid jet. The time series in Fig.~\ref{fig:cav_dyn_jet2} visualise the cavitating flow in the nozzle together with the liquid jet injected into air. For $\sigma = 1.19$, no strong effect on the jet is present. However, the jet opening angle is increased on the cavitating side and the jet surface is more disturbed at this side as well. On the other side, for $\sigma = 0.84$ cavitation and partial gas entrainment have a strong impact on the jet appearance,  especially close to the nozzle outlet. In the depicted time series partial gas entrainment leads to a bulging of the jet surface. At $t=0.76\;\si{ms}$ the jet surface starts to bulge where the gas is sucked in and the spray angle near the outlet is suddenly strongly increased. Afterwards this bulge grows ($t=0.81\;\si{ms}$) and is convected downstream, where it bursts away and finally further detaches. This phenomenon was also described by \citet{Orley:2015kt,Edelbauer:2017fx}.

\begin{figure}[!ht]
\centering
 \subfigure[]{\includegraphics[height=5.5cm]{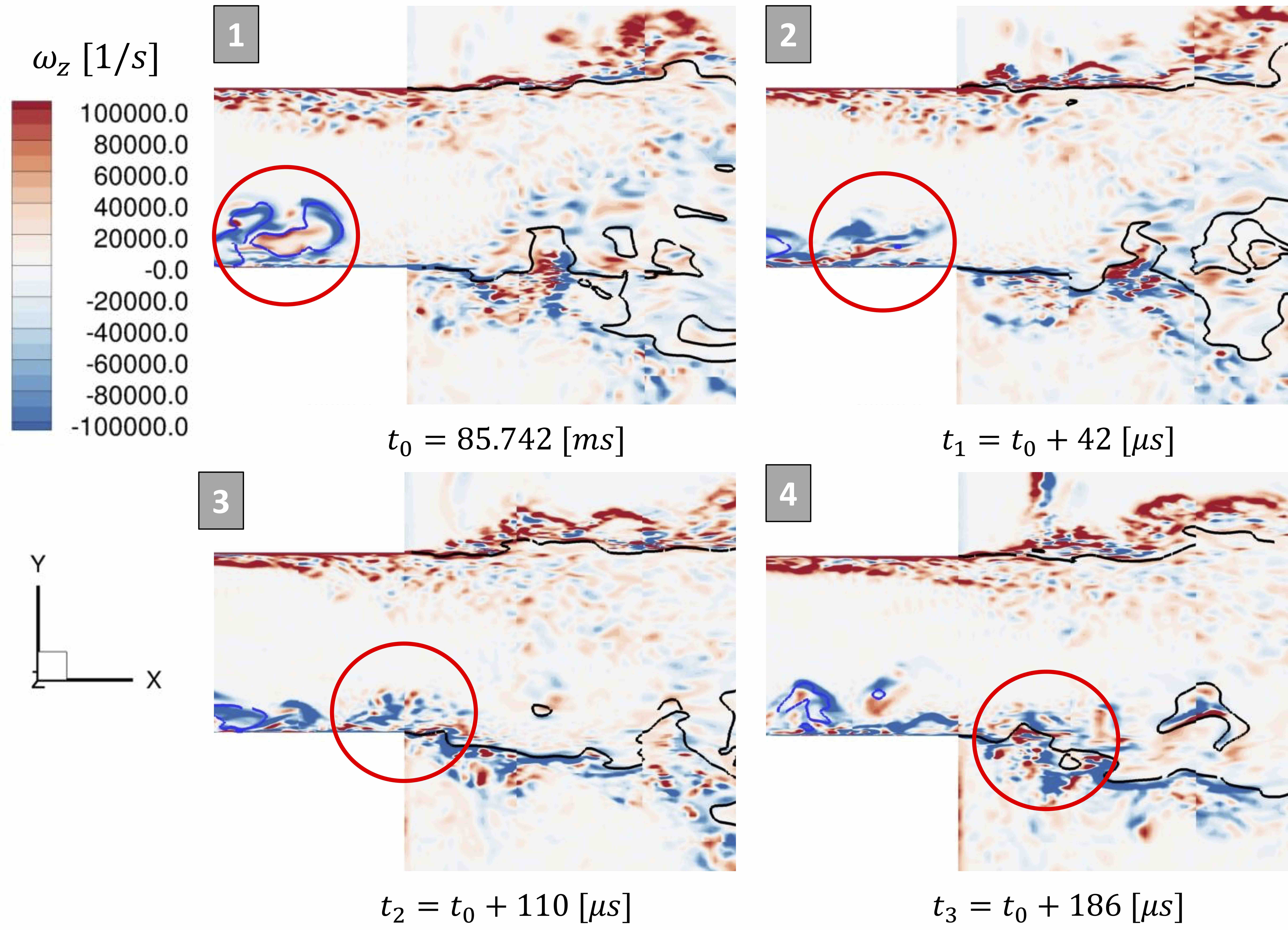}}
 \subfigure[]{\includegraphics[height=5.5cm]{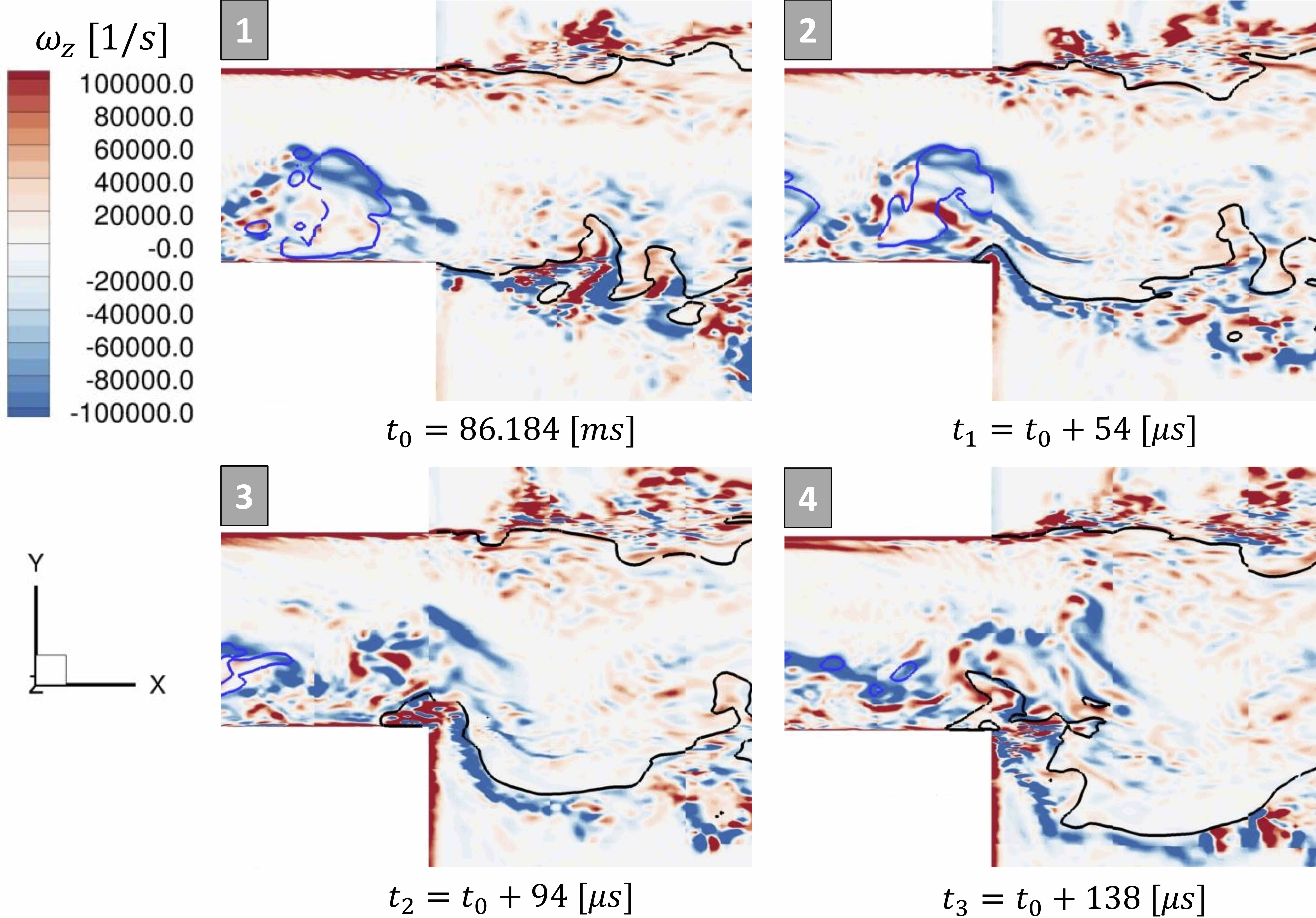}}
\caption{Vorticity after a collapse event (a) and after gas entrainment (b) for $\sigma = 0.84$;  iso-lines: vapor (blue,10$\%$) gas (black,75$\%$).(Color online) }
\label{fig:vorticity}
\end{figure}

\begin{figure}[!htb]
  \centering
  \includegraphics[width=0.99\linewidth]{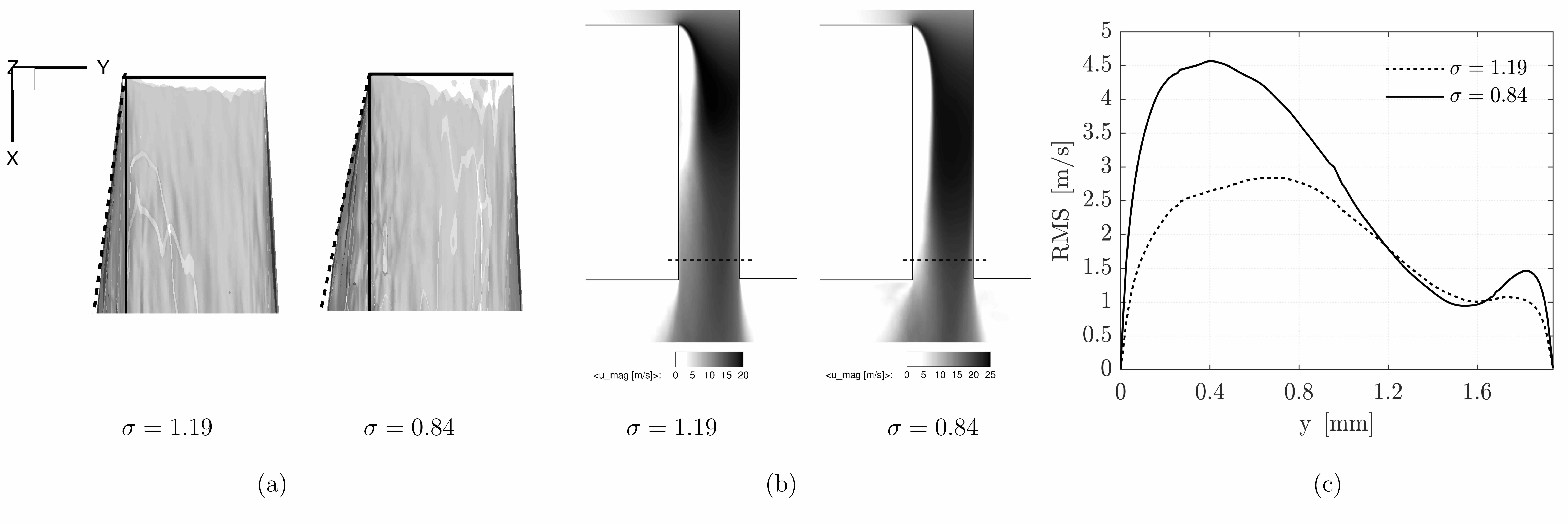}
  \caption{(a) Averaged jet surface $75\%$. (b) Averaged velocity magnitude on the mid plane, contour line at $10\%$ vapor and marked position of the evaluation of the fluctuations. (c) Fluctuations in lateral direction $0.5\;\si{mm}$ before the nozzle outlet (nozzle length $8\;\si{mm}$). All data are averaged over $4\;\si{ms}$.}
\label{fig:angle}
\end{figure}

In experiments \citep{Sou:2007jd} and numerical investigations \citep{Orley:2015kt}, collapse induced fluctuations have been found to enhance primary jet break-up. Besides the increased fluctuations, \citet{Orley:2015kt} proposed two additional mechanisms: collapses inside the jet close to the surface and gas entrainment into the nozzle. In the following we discuss and evaluate the influence of gas entrainment and collapse events on the jet characteristic. Fig.~\ref{fig:vorticity} compares vorticity induced by collapse events and by gas entrainment. In Fig.~\ref{fig:vorticity}~(a) a vapor structure collapses close to the nozzle outlet and induces a local increase of turbulent fluctuations, which are convected outside and disturb the jet surface. On the other hand, gas entrainment is visualized in Fig.~\ref{fig:vorticity}~(b), where gas from the outlet area enters the nozzle and leads to a higher vorticity than in (a) with larger spatial extent and subsequently to a strong increase of the near-nozzle spray angle. Thus, we conclude that in the investigated configuration partial gas entrainment may have a more significant influence on the jet characteristic than collapse events. 
The temporally averaged data, see Fig.~\ref{fig:angle}, also confirm an  increased mean jet angle in the super-cavitating case ($\sigma = 0.84 $). For $\sigma = 1.19$ the average jet opening angle on the cavitation side is \ang{6.5} and for $\sigma = 0.84 $ it is \ang{12.4}. Both values are in good agreement with experimental data \citep{Sou:2007jd, Stanley:2011gr}, although at $\sigma = 1.19$ our value is slightly higher than in the experiments, where it was found to lie between $\ang{4}-\ang{5}$. 

According to experimental investigations by \citet{Sou:2007jd}, velocities  and fluctuations in lateral direction are among the key parameters for an enhanced jet break-up. Both can lead to a higher disturbance of the jet surface in lateral direction.  Consequently, we now consider the average flow field in  Fig.~\ref{fig:angle}~(b) and the lateral velocity fluctuations \SI{0.5}{mm} before the nozzle outlet in Fig.~\ref{fig:angle}~(c).  For $\sigma = 1.19$, the flow reattaches approximately at half the nozzle length, whereas for $\sigma = 0.84$ reattachment takes place close to the outlet. The lateral velocity fluctuations are higher on the side of the nozzle where flow detachment and cavitation occurs. However, at $\sigma = 0.84$ they are significantly increased compared to $\sigma = 1.19$. A higher $y$ values ($1-1.6\;\si{mm}$), where no influence by cavitation or flow detachment is present, the fluctuations are on a similar level. At the opposite wall they increase again and are higher for $\sigma = 0.84$.  These observations confirm the experimental findings by \citet{Sou:2007jd}. 
\nomenclature[G]{$\omega$}{Vorticity \nomunit{[\si{1/s}]}}

\section{Conclusions}
\label{sec:conclusion}

We have performed implicit LES at two operating points covering the regimes of \textit{developing cavitation} and \textit{super cavitation} using a fully compressible multi-phase model. The set-up is adapted from an experiment where tap water is discharged through a large-scale nozzle into air. Our simulation results have been validated with experimental flow field data. 

Based on the vapor content, the shedding frequencies of the partial cavitation were determined. Consistently with theory, the dominant frequency shifts to lower values for increased cavitation. The Strouhal numbers are in good agreement with values from literature. With spectral analysis we were able to confirm oscillations with the same frequency for the mass flows at the nozzle inlet and outlet as well as for the integral vapor volume in the nozzle. Discharge coefficients were calculated and match the experimental ones. Additionally, we found gas entrainment to influence strongly the mass flow at the nozzle outlet. Gas entrainment can lead to a temporary drop of the mass flow by nearly 20 $\%$. This can be crucial in injection processes, if their efficiency is governed by the mass flow. 

Our computation reproduces the interaction of the cavitating nozzle flow with the emanating jet. 
In the case of \textit{super cavitation}, detached vapor structures can reach the nozzle outlet, leading to partial entrainment of gas from the outflow region into the nozzle. Due to the similar physical properties of vapor and gas, this mechanism is hard to capture experimentally. Our multi-component computation, however, allows to a deeper insight into this phenomenon and its effects. Visualization of the numerical results revealed partial gas entrainment as one of the driving mechanisms for a widening of the jet. The other main factor is the presence of higher velocity fluctuations close to the nozzle outlet in the case of strong cavitation, which are induced by collapse events. Time averaged data confirm that increased lateral fluctuations close to the nozzle outlet correlate with an increased averaged jet angle. 

The performed numerical analyses serve as a basis for further investigations with higher injection pressures and realistic geometries.

\section*{Acknowledgments}
The authors are grateful to acknowledge the Gauss Centre for Supercomputing e.V. for providing computing time on the GCS Supercomputers SuperMUC at Leibniz Supercomputing Centre (LRZ, www.lrz.de). 

\begin{mdframed}
\printnomenclature
\end{mdframed}

\pagebreak 

\bibliography{main}
\bibliographystyle{elsarticle-harv}
\biboptions{authoryear}
\end{document}